\documentclass[aps,preprint,groupedaddress,showpacs,showkeys]{revtex4}
\usepackage{bm,amsmath}
\usepackage{graphicx}
\usepackage[font={small}]{caption}
\usepackage[caption=false]{subfig}
\usepackage{float}

\begin{document}

\title{Instability in Reaction-Superdiffusion Systems}

\author{Reza Torabi}
\email{rezatorabi@aut.ac.ir}

\author{Zahra Rezaei}
\email{z.rezaei@aut.ac.ir}
\affiliation{Department of Physics, Tafresh University, Tafresh 39518 79611, Iran}

\begin{abstract}
We study the effect of superdiffusion on the instability in
reaction-diffusion systems. It is shown that reaction-superdiffusion
systems close to a Turing instability are equivalent to a
time-dependent Ginzburg-Landau model and the corresponding free
energy is introduced. This generalized free energy which depends on
the superdiffusion exponent governs the stability, dynamics and the
fluctuations of reaction-superdiffusion systems near the Turing
bifurcation. In addition, we show that for a general n-component
reaction-superdiffusion system, a fractional complex Ginzburg-Landau
equation emerges as the amplitude equation near a Hopf instability. Numerical simulations of this equation are carried out to illustrate the effect of superdiffusion on spatio-temporal patterns. Finally the effect of superdiffusion on the instability in
Brusselator model, as a special case of reaction-diffusion
systems, is studied. In general superdiffusion introduces a new
parameter that changes the behavior of the system near the
instability.

\end{abstract}

 \maketitle

\maketitle

\section{Introduction}

Self-organized phenomena are ubiquitous in nature specially in
living systems. They occur in open systems out of equilibrium and
they have attracted the attention of scientists in different fields
of science. In the study of self-organized phenomena,
reaction-diffusion systems are extensively used
\cite{Nicolis,Kerner,Buceta,Mimura}. They are
sets of coupled partial differential equations which include
diffusion terms. Reaction-diffusion systems are useful in many fields such as biology
\cite{Harrison,Meinhardt,Murray1,Murray2,Chaplain,Sherratt1,Sherratt2,Gatenby,Kondo,Shin},
ecology \cite{Holmes,Skellam}, neuroscience \cite{Liang}, physics
\cite{Astrov,Arecchi,Staliunas}, chemistry \cite{Castets} and geology \cite{Heureux}. They
have rich dynamics and can produce spatio-temporal patterns
\cite{Walgraef,Haken,Murray3,Cross1,Cross2,Aranson} including
traveling waves, kinks, vortices, domain walls, solitons, as well as
hexagonal and stripe patterns.

Reaction diffusion systems were first proposed by Alan Turing in the
study of morphogenesis \cite{Turing}. Actually, Turing noticed that
adding a diffusion term to a reaction system can derive the system
to instability and plays an important role in the formation of
spatio-temporal patterns out of equilibrium. There are two possible
kinds of instability called Turing and Hopf. In fact, in a
reaction-diffusion system, a steady state can experience a
transition to an oscillating or a patterned state via a Hopf or
Turing instability, respectively \cite{Kuramoto}. The characteristic
feature of most of the studied reaction-diffusion systems is that
the diffusion is normal. However, experimental evidences show that
anomalous diffusion arises more frequently in nature
\cite{Bouchaud,Haus,Metzler1,Metzler2,Sokolov}. This fact motivated
us to consider anomalous diffusion in reaction-diffusion systems in
this article.

In a normal diffusion the mean square displacement of a typical
particle of the system grows linearly with time, $\langle
x^2(t)\rangle \propto t$. Anomalous diffusion, on the other hand, is
a diffusion process that does not obey this linear relation. In the
most cases they satisfy a power law scaling relation, $\langle
x^2(t)\rangle\propto t^\gamma$, which is present in variety of
different systems. $\gamma$ is called anomalous diffusion exponent
and for $\gamma=1$ we obtain the case of normal diffusion.
$1<\gamma<2$, $0<\gamma<1$ and $\gamma=2$ correspond to a Levy superdiffusion, a subdiffusion and a ballistic diffusion, respectively \cite{Metzler1}. Both types
of anomalous diffusion processes play important roles in various
phenomena
\cite{Wachsmuth,Weiss,Drazer,Amblard,Carreras,Hansen,Solomon,Manandhar,Sancho,Angelico,Viswanathan2,Sims,Scher}.
For instance, subdiffusion often occurs in gels (especially bio-gels
\cite{Wachsmuth,Weiss}), porous media \cite{Drazer}, and polymers
\cite{Amblard}. Levy superdiffusion is typical of some processes in
plasmas and turbulent flows \cite{Carreras,Hansen,Solomon}, surface
diffusion \cite{Manandhar,Sancho,Angelico}, animals hunting
(especially for ocean predators and birds) \cite{Viswanathan2,Sims}
and charge carrier transfer in semiconductors \cite{Scher}.

Although different aspects of anomalous diffusion as well as the
properties of reaction-diffusion processes have been extensively
studied separately, reaction-diffusion systems characterized by
anomalous diffusion have been the subject of a limited number of
studies
\cite{Zhang,Golovin,Gambino,Gafiychuk,Henry,Henry2,Langlands,Weiss2,Nec,Tzou,Tzou2}. Addressing this shortcoming on one hand and the importance of
recognition and control of instability in far from equilibrium
systems on the other hand motivated us to study the instability of
reaction-diffusion systems in the presence of anomalous superdiffusion.

Turing pattern formation in the Brusselator model with superdiffusion has been studied in \cite{Golovin}. The authors have focused on the pattern selection in the formation of hexagons and stripes and have compared the case of normal and superdiffusion. Turing pattern formation has also been investigated in a typical activator-inhibitor system \cite{Zhang} with a reaction term that has been first used to describe the chlorite-iodine-molonic acid \cite{Dufiet}. It was found that the wave vector of patterns changes with the superdiffusion exponent which leads to a different size for Turing patterns. Nonlinear dynamics of an activator-inhibitor system with superdiffusion near the Hopf instability has been studied in \cite{Nec}. It was shown that a fractional complex Ginzburg-Landau equation (FCGLE) governs the amplitude of critical mode in the vicinity of Hopf instability for two-component reaction-superdiffusion systems. Apart from mentioned studies, spatio-temporal patterns near a codimension-2 Turing-Hopf point, where Turing and Hopf instability thresholds coincide, have been considered in \cite{Tzou} for a one dimensional superdiffusive Brusselator model and the long-wave stability of these patterns has been analyzed in \cite{Tzou2}.

Instability in a reaction-diffusion system is an example of non-equilibrium phase transitions. On the other hand in the vicinity of critical points fluctuations play an important role and the systems exhibit universal behavior. This motivated us to study the behavior of reaction-superdiffusion systems at the onset of instabilities and investigate the spectrum of fluctuations in the presence of superdiffusion. As the main part of our study we will show that any reaction-superdiffusion system in the vicinity of a Turing instability is equivalent to a time-dependent Ginzburg-Landau model. We will introduce the corresponding free energy which depends on the superdiffusion exponent. This free energy governs the instability, dynamics and the fluctuations of the system. In addition in the case of Hopf instability we generalize the two-component system of \cite{Nec} to a n-component reaction-superdiffusion system. Utilizing the reductive perturbation method \cite{Kuramoto} we show that a fractional complex Ginzburg-Landau equation governs the amplitude of the critical mode for a general n-component system, too. The solutions of FCGLE in two dimensions display a very rich spectrum of dynamical behavior. So we present numerical simulation of FCGLE in two dimensions to illustrate the effect of superdiffusion on spatio-temporal patterns near the Hopf instability. Finally we apply these general instability considerations to the Brusselator model in the presence of superdiffusion. Brusselator is a typical example of a reaction-diffusion system and is one of the most common non-linear chemical systems \cite{Golovin,Reichl,Walgraef2}.

This paper is organized as follows. In section II we review how
superdiffusion can be considered in reaction-diffusion systems using
the powerful tool of fractional calculus. The behavior of the
reaction-superdiffusion system close to the Hopf and Turing
instabilities is respectively investigated in sections III and IV.
In section V the general instability considerations of the previous
sections are applied to the Brusselator model. Finally conclusions
and discussions are presented in section VI.

\section{Brief overview of reaction-superdiffusion systems}

In this section we are going to consider anomalous diffusion in a
reaction-diffusion system. In the introduction we presented a
microscopic definition for the anomalous diffusion while
reaction-diffusion systems are differential equations governing
macroscopic quantities. Therefore we have to first introduce a
macroscopic representation for the anomalous diffusion. To do this
one can start with the microscopic point of view and then obtain a
macroscopic equation for the anomalous diffusion in the continuum
limit.

Suppose a normal diffusion. From a microscopic point of view, a normal
diffusion is described by random motion of a particle with equal
length of steps and equal waiting times between successive steps.
This is the Brownian motion that in the continuum limit leads to the
differential equation governing the normal diffusion process
\cite{Metzler1}. However, both waiting time between successive jumps
and the length of the steps may not be equal and can be extracted
from continuous probability distribution functions. This is called
the continuous random walk (CTRW) model which is used to describe
anomalous diffusion \cite{Metzler1,Metzler2,Sokolov}. In the
subdiffusion process, due to the particle sticking and trapping, the
waiting time probability distribution function is a heavy tailed
function while in the case of Levy superdiffusion the length of
steps obey such a heavy tailed function. When a particle experiences
Levy flight instead of a Brownian motion, large jumps occur more
frequently than the case of Brownian motion. Using CTRW approach
accompanied by considering the power-law distribution function
\begin{equation}\label{powerlaw}
P(x)\propto |x|^{-(1+\alpha)},
\end{equation}
for steps, results in the one dimensional fractional diffusion equation
\begin{equation} \nonumber
\frac{\partial n}{\partial t}=D_\alpha \frac{\partial^\alpha
n}{\partial x^\alpha},
\end{equation}
in the continuum limit \cite{Metzler1} where $D_\alpha$ is the
generalized diffusion coefficient. $\alpha$ is the fractional order
of the derivative that relates to $\gamma$ via $\alpha=2/\gamma$
where $\gamma$ appeared in the previously mentioned microscopic
equation for mean square displacement of a particle in anomalous
diffusion, $\langle x^2(t)\rangle\propto t^\gamma$. $n$ is a typical
concentration in the system and $\frac{\partial^\alpha}{\partial
x^\alpha}$ is the Weyl fractional operator $(1 < \alpha < 2)$ that
is defined as
\begin{equation} \nonumber
 \frac{\partial^\alpha
n}{\partial
x^\alpha}=-\frac{1}{2\cos(\pi\alpha/2)}(\partial_+^\alpha
n+\partial_-^\alpha n),
\end{equation}
\begin{equation} \nonumber
\partial_+^\alpha
n=\frac{1}{\Gamma(2-\alpha)}\frac{d^2}{dx^2}\int_{-\infty}^x\frac{n(q,t)}{(x-q)^{\alpha-1}}
dq,
\end{equation}
\begin{equation} \nonumber
\partial_-^\alpha
n=\frac{1}{\Gamma(2-\alpha)}\frac{d^2}{dx^2}\int^{\infty}_x\frac{n(q,t)}{(q-x)^{\alpha-1}}
dq,
\end{equation}
where $\Gamma$ stands for the Gamma function \cite{Metzler1,Zhang,Golovin}. In higher dimensions,
$\frac{\partial^\alpha }{\partial x^\alpha}$ operator is replaced by
$\nabla^\alpha$, defined by its action in Fourier space,
$\mathcal{F}(\nabla^\alpha n) = -|{\bf k}|^\alpha \mathcal{F}(n)$. Note that $\alpha$ in the probability
distribution of jumps, \eqref{powerlaw}, equals the order of the
fractional derivative. Another interpretation for the index $\alpha$
comes from considerations of fractals and self-similarity. The path
of a Brownian particle in space traces out a random fractal of
dimension two, while a Levy particle draws a fractal of dimension
$\alpha$.

A general n-component reaction-diffusion system is
described by the following differential equation \cite{Kuramoto}
\[
\frac{\partial {\bf X}}{\partial t}={\bf F}({\bf
    X};\mu)+\mathcal{D} \nabla^2 {\bf X},
\]
where ${\bf X}$ and ${\bf F}$ are $n$-dimensional real vectors,
$\mu$ is the bifurcation parameter and $\mathcal{D}$ is a diagonal matrix of diffusion coefficients.
Therefore according to our discussion in this section, a general n-component reaction-superdiffusion system can be given by
\begin{equation} \label{Reacdif}
\frac{\partial {\bf X}}{\partial t}={\bf F}({\bf
    X};\mu)+\mathcal{D}_\alpha \nabla^\alpha {\bf X},
\end{equation}
where $\mathcal{D}_\alpha$ is a
diagonal matrix of generalized diffusion coefficients.
The characteristic feature of a reaction-superdiffusion system is
that the fractional derivative introduces a new parameter, $\alpha$,
that changes the properties of the solution.

\section{Hopf instability in reaction-superdiffusion systems}

In this section we will study the behavior of a
reaction-superdiffusion system in the vicinity of a Hopf
bifurcation. Consider the general differential equation
\eqref{Reacdif} governing a reaction-superdiffusion system. As $\mu$
varies the system may move from a steady state to a time-periodic
state (limit cycle) near a Hopf bifurcation. Close to criticality,
we are left with a couple of relevant dynamical variables whose time
scales are distinguishably slower than the other dynamical
variables, so that the latter can be eliminated adiabatically using
the rescaled spacetime coordinate. This technique is called
reductive perturbation method. As a result, \eqref{Reacdif} is
contracted to a very simple universal equation. Here, in subsection
A, we show that, in the case of a Hopf bifurcation, it is a
fractional complex Ginzburg-Landau equation. Then the numerical
study of FCGLE is presented in subsection B.

\subsection{Formal Approach}

\begin{figure}
\begin{center}
\includegraphics[scale=0.4]{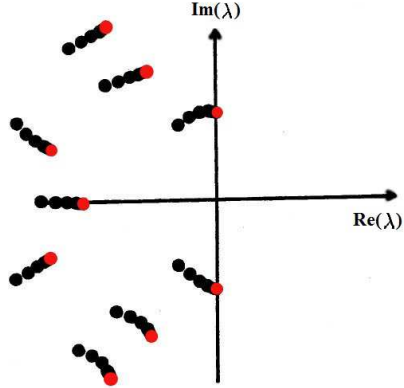}
\caption{ Schematic distribution of the eigenvalues. The eigenvalues
of uniform and nonuniform modes have been shown in red and black,
respectively \cite{Kuramoto}.}\label{figlambda}
\end{center}
\end{figure}

Let ${\bf X}_0(\mu)$ be the steady solution of \eqref{Reacdif},
${\bf F}({\bf X}_0;\mu)=0$. Taylor series expansion of
\eqref{Reacdif} about the steady state, ${\bf u}({\bf r},t)={\bf
X}-{\bf X}_0$, leads to
\begin{equation} \label{Taylor}
\frac{\partial {\bf u}}{\partial t}=({\bf L}+\mathcal{D}_\alpha
\nabla^\alpha) {\bf u}+{\bf Muu}+{\bf Nuuu}+...,
\end{equation}
where ${\bf L}$ is the Jacobian matrix whose $ij$th element is
$L_{ij}=\partial F_i({\bf X}_0)/\partial X_j$. ${\bf Muu}$ and ${\bf
Nuuu}$ are nonlinear terms that denote vectors (summation
convention is used)
\begin{equation} \label{MUU}
{\bf Muu}=\frac{1}{2}\frac{\partial^2 {\bf F}({\bf X}_0)}{\partial
X_j \partial X_k}u_j u_k,
\end{equation}
\begin{equation} \label{NUUU}
{\bf Nuuu}=\frac{1}{6}\frac{\partial^3 {\bf F}({\bf X}_0)}{\partial
X_j \partial X_k \partial X_l}u_j u_k u_l.
\end{equation}
Note that the expansion coefficients, which are symbolically
expressed by ${\bf M},{\bf N},...$ generally depend on $\mu$ at
least through ${\bf X}_0(\mu)$. Assume that up to $\mu_c=0$, ${\bf
X}_0$ is stable against small perturbations, while it loses stability for
$\mu>0$. For a while, let's forget about the special degrees of freedom
coming from the fractional derivative in \eqref{Reacdif}. The
stability of ${\bf X}_0$ depends on the configuration of eigenvalues
$\lambda$ given by
\begin{equation} \nonumber
{\bf L}{\bf u}=\lambda {\bf u}. \label{eigen}
\end{equation}
A hypothetical configuration of eigenvalues is plotted in Fig. \ref{figlambda}.
Allowing, now, the diffusion term to be present causes the special
modes to come into play. The linearized equation, in this case will
be
\begin{equation} \nonumber
\frac{\partial {\bf u}}{\partial t}=( {\bf
L}+\mathcal{D}_\alpha\nabla^\alpha){\bf u}.
\end{equation}
Putting the normal mode solution of the form ${\bf u}={\bf V}_{\bf
k} e^{\lambda t} e^{i{\bf k}.{\bf r}}$ in the above equation, one
can easily find that in the presence of diffusion term a bunch of
eigenvalues (corresponding to non-uniform modes) appear near each
eigenvalue of ${\bf L}$ (Fig. \ref{figlambda}). Also, the distance
between two neighboring eigenvalues in the same branch is found to
be of the order $\xi^{-\alpha}$ where $\xi\sim 1/k$ is a measure of
the length of nonuniform modes and $k=|{\bf k}|$. This means that in
each branch nonuniform modes with small wave vectors has eigenvalues
close to the uniform mode. Therefore, these modes have comparable
growth and decay rates with the uniform mode and affect the behavior
of the system near the critical point ($\mu\rightarrow 0$). So, we
have to take them into account.

Near the criticality, ${\bf L}$, ${\bf M}$, ${\bf N}$, eigenvectors
and eigenvalues can be expanded in powers of $\mu$ as
\begin{eqnarray}\label{expand}
&{\bf L}={\bf L^{(0)}}+\mu {\bf L^{(1)}}+\mu^2 {\bf L^{(2)}}+...,\nonumber \\
&{\bf M}={\bf M^{(0)}}+\mu {\bf M^{(1)}}+\mu^2 {\bf M^{(2)}}+... ,\nonumber \\
&{\bf N}={\bf N^{(0)}}+\mu {\bf N^{(1)}}+\mu^2 {\bf N^{(2)}}+... ,\nonumber \\
&{\bf u}=\mu^{\frac{1}{2}}{\bf u^{(1)}}+\mu {\bf u^{(2)}}+\mu^{\frac{3}{2}} {\bf u^{(3)}}+... ,\nonumber \\
&\lambda=\lambda^{(0)}+\mu \lambda^{(1)}+\mu^2 \lambda^{(2)}+...,\nonumber \\
\end{eqnarray}
where $\lambda^{(0)}=\pm i\omega_0$ and $\lambda^{(i)}=\sigma_i+
i\omega_i$. We must get inside the neighborhood of the critical mode
and we do this by rescaling the space and time variables. Since
$\lambda$ has real part of order $\mu$ and the characteristic time
is equal to the inverse of the real part of $\lambda$, time is
naturally rescaled according to
\[
\tau=|\mu| t.
\]
However, rescaling the space requires more details. In fact we
should define a length scale for which nonuniform modes become
important in the dynamics of the system. For this purpose, note that
the characteristic time scale of critical modes is $\tau_0\sim
\mu^{-1}$ \cite{Kuramoto}. However, for the slowest nonuniform modes
it roughly is $\tau_{\nu}\sim(\mu+\xi^{-\alpha})^{-1}$ where we have
ignored $D_{\alpha}$ in the scaling argument. These two
characteristic times are of the same order if
$|\mu|\sim\xi^{-\alpha}$. Therefore, those nonuniform modes whose
wavelengths are greater than $\xi>|\mu|^{-\frac{1}{\alpha}}$ play a
role in the long time behavior of the system. This suggests us to
introduce a scaled coordinate $s$ defined by
\[
s=|\mu|^\frac{1}{\alpha}r=\varepsilon^{\frac{2}{\alpha}}r,
\]
where $\varepsilon^2 \chi\equiv\mu$ and $\chi=sgn(\mu)$. Based on the
above discussion, ${\bf u}$ is regarded as a function of $t$,
$\tau$, and $s$. This means that we are dealing with the long time,
long wavelength modes in their natural variables  $\tau$ and $s$,
and reserving $t$ for the overall periodic motion (limit cycle).
Also note that $\nabla\rightarrow \varepsilon^{2/\alpha}\nabla_{\bf
s}$, so, $\nabla^\alpha\rightarrow \varepsilon^{2}\nabla^\alpha_{\bf
s}$. Substitution of \eqref{expand} into \eqref{Taylor} and equating
coefficients of different powers of $\varepsilon$, yields to a set
of equations in the form of
\begin{equation}\label{homo}
(\frac{\partial}{\partial t}-{\bf L^{(0)}}){\bf
u^{(\boldsymbol\nu)}}={\bf B_{\boldsymbol\nu}}, \;\;\; \nu=1,2,...,
\end{equation}
where the first three $\bf B$'s are
\begin{eqnarray}
&{\bf B_1}={\bf 0},\nonumber\\
&{\bf B_2}={\bf M^{(0)}} {\bf u^{(1)}}{\bf u^{(1)}},\nonumber\\
&{\bf B_3}= -(\frac{\partial}{\partial \tau}-\chi {\bf
L^{(1)}}-\mathcal{D}_\alpha\nabla^\alpha_{\bf s}){\bf u^{(1)}}+2{\bf
M^{(0)} u^{(1)} u^{(2)}}+{\bf N^{(0)}
u^{(1)}u^{(1)}u^{(1)}}.\nonumber
\end{eqnarray}
Note that the term with fractional derivative contributes to the coefficient of $\varepsilon^3$.

There is a solvability condition for the set of equations \eqref{homo} \cite{Kuramoto}, which for $\nu=1$ leads to
\[
{\bf u^{(1)}}(t,\tau,s)=W(\tau,s){\bf U_R}e^{i\omega_0 t}+c.c,
\]
where $c.c.$ stands for the complex conjugate, ${\bf U_R}$ is the
right eigenvector of ${\bf L^{(0)}}$, and $W(\tau,s)$ is a complex
amplitude to be determined. The right eigenvector and left
eigenvector $({\bf U_L})$ are normalized in such a way that ${\bf
U_L}{\bf U_R}=1$. The solvability condition for $\nu=2$ gives rise
to an expression for ${\bf u^{(2)}}$
\begin{equation} \nonumber
{\bf u^{(2)}}={\bf V_+} W^2 e^{2i\omega_0 t}+{\bf V_-} \overline{W}^2
e^{-2i\omega_0 t}+{\bf V_0} |W|^2,
\end{equation}
where
\begin{equation} \nonumber
{\bf V_+}={\bf \overline{V}}_-=-({\bf L^{(0)}}-2i\omega_0 {\bf
    I})^{-1} {\bf M^{(0)} U_R U_R},
\end{equation}
\begin{equation} \nonumber
{\bf V_0}=-2{\bf L^{(0)}}^{-1} {\bf M^{(0)} U_R \overline{U}_R}.
\end{equation}
The bar stands for complex conjugate, ${\bf I}$ is the identity
matrix and the operator ${\bf M}$ can be read off from \eqref{MUU}.
Putting ${\bf u^{(1)}}$ and ${\bf u^{(2)}}$ into the solvability
condition for $\nu=3$ results in the equation governing the
amplitude $W$
\begin{equation}\label{CGLEF}
\frac{\partial W}{\partial \tau}=\chi \lambda^{(1)}
W+d\nabla^\alpha_{\bf s} W-g|W|^2W,
\end{equation}
where $\lambda^{(1)}$, $d$ and $g$ are generally complex numbers that are given by
\begin{eqnarray} \label{CGLEcoeff} %
\lambda^{(1)} &=& {\bf U_L} {\bf L^{(1)}} {\bf U_R}, \cr %
d &=& d_r+id_i={\bf
    U_L}\mathcal{D}_\alpha {\bf U_R}, \cr %
g &\equiv & g_r+ig_i=4{\bf U_L M^{(0)} U_R}{\bf L^{(0)}}^{-1} {\bf
M^{(0)} U_R
\overline{U}_R}\cr %
&+&2{\bf U_L M^{(0)} \overline{U}_R}({\bf L^{(0)}}-2i\omega_0 {\bf
I})^{-1} {\bf M^{(0)} U_R U_R} -3{\bf U_L N^{(0)} U_R U_R
\overline{U}_R}.
\end{eqnarray}
\eqref{CGLEF} is a fractional Ginzburg-Landau equation and has
non-trivial oscillatory solutions, in the absence of diffusive term,
provided that $g_r$ and $\chi$ have the same sign
\cite{Kuramoto,Moralesa}. These solutions are stable in the case of
$\chi>0$, supercritical Hopf bifurcation, and unstable for $\chi<0$,
subcritical Hopf bifurcation. So the stability condition implies the
simultaneous establishment of the conditions $g_r >0$ and $\chi >0$.
We implicitly assume that these conditions are fulfilled here. With
a redefinition as follows
\[
r'= ({\sigma_1/d_r})^{1/\alpha} s,\;\;\; t'=\sigma_1 \tau, \;\;\;
W'= \sqrt{g_r/\sigma_1}e^{-i\omega_1 \tau} W,
\]
FCGLE above criticality can be written in a more convenient form
(dropping the primes)
\begin{equation}\label{CGLERS}
\frac{\partial W}{\partial t}= W+(1+ic_1)\nabla^\alpha
W-(1+ic_2)|W|^2W,
\end{equation}
where $c_1=d_i/d_r$ and $c_2=g_i/g_r$. It is obvious from \eqref{CGLERS} that superdiffusion
can change the properties of the solution by changing the order of
derivative as well as the parameter $c_1$.

To study \eqref{CGLERS} in more details, note that the first term of
the r.h.s is related to the linear instability mechanism which leads
to oscillations. The second term accounts for diffusion and
dispersion while the third one is a cubic non-linear term. The
competition between these three terms results in different regimes.
There are two interesting limits for \eqref{CGLERS} which are worth
mentioning. For $c_1, c_2\rightarrow 0$, \eqref{CGLERS} reduces to a
time-dependent fractional real Ginzburg-Landau equation (FRGLE) and
as $c_1, c_2\rightarrow \infty$, one obtains a fractional nonlinear
Schr\"{o}dinger equation.

Analytical and numerical solutions of FCGLE in one dimension and
also some aspects of numerical simulations in two dimensions have
been presented in \cite{Nec}. Since the solutions of FCGLE in two
dimensions display a very rich spectrum of dynamical behavior we
were encouraged to present a comprehensive numerical study of these
solutions in two dimensions in the next subsection. This enables us
to observe the effect of fractional order, $\alpha$, on the
spatio-temporal patterns.

\subsection{Numerical Simulations}

As was mentioned, FCGLE possesses a rich dynamics in two dimensions
and such as its special case, CGLE \cite{Chate}, has three different
regimes. There are two kinds of disordered regimes called {\it phase
turbulence } and {\it defect turbulence}, depending on whether they
exhibit defects or not. In phase turbulence regime the amplitude of
the field $W$ never reaches zero while this amplitude becomes zero
at some points for defect (amplitude) turbulence regime. Apart from
these two disordered regimes there are {\it frozen states} those in
which $|W|$ is stationary in time. Cellular structures and spiral
patterns can be observed in this regime.

As we saw in Eq. \eqref{CGLERS}, superdiffusion changes both the
order of spatial derivative and the parameter $c_1$. On the other
hand, variation of any of the parameters in CGLE can affect the
solutions \cite{Chate}. Here we are interested in the effect of
fractional derivative on the solutions of FCGLE, so we neglect the
dependence of $c_1$ on $\alpha$.

For solving FCGLE we use a pseudospectral method to perform
numerical computations in Fourier space. The numerical simulation is
based on the method of exponential time differencing (ETD2)
\cite{Cox}. Small amplitude random initial data about $W=0$ and
periodic boundary conditions are used.

\begin{figure}[H]
\begin{center}
       \subfloat[] {\includegraphics[width=1.8 in]{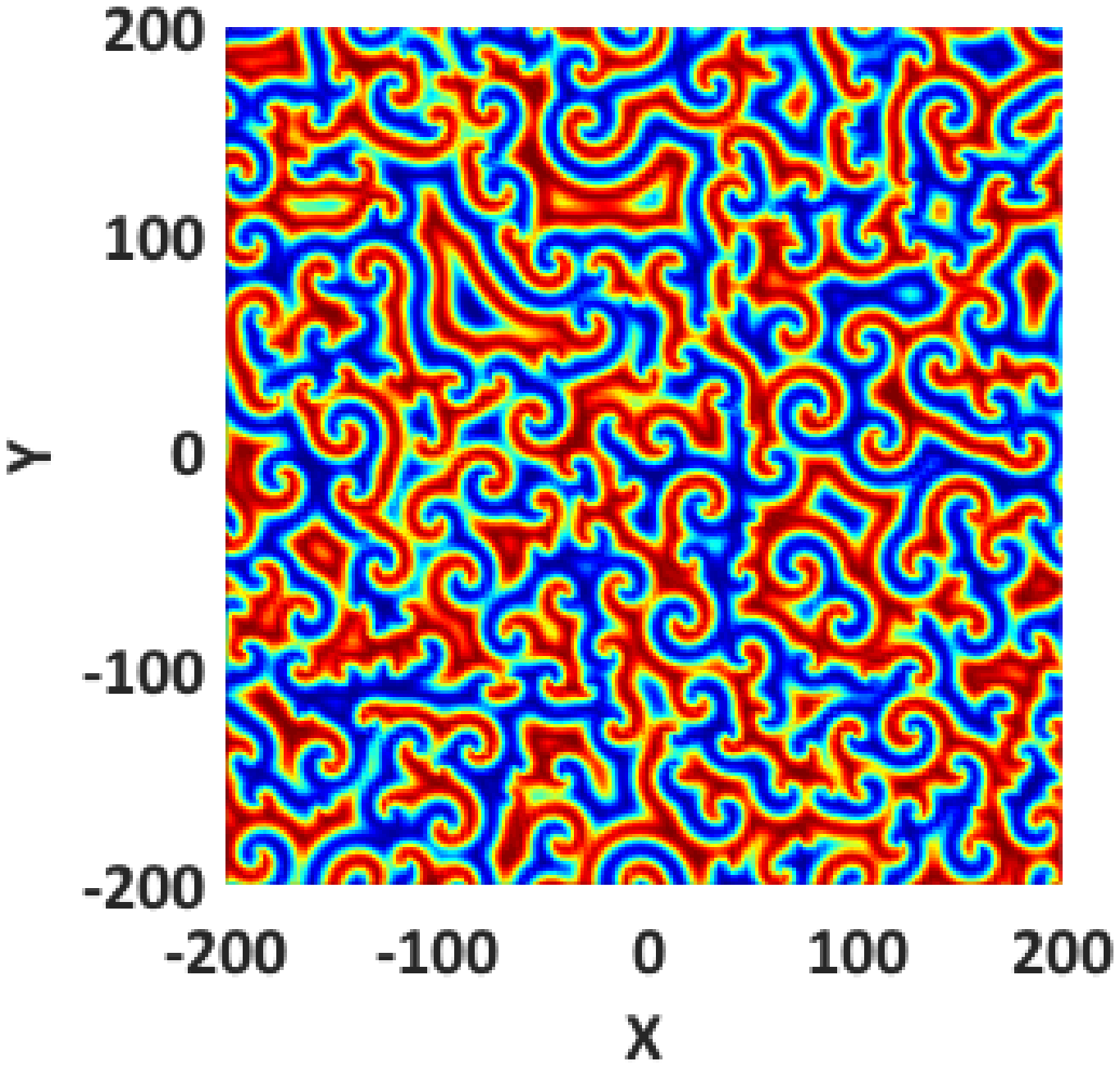}}
       \subfloat[] {\includegraphics[width=1.8 in]{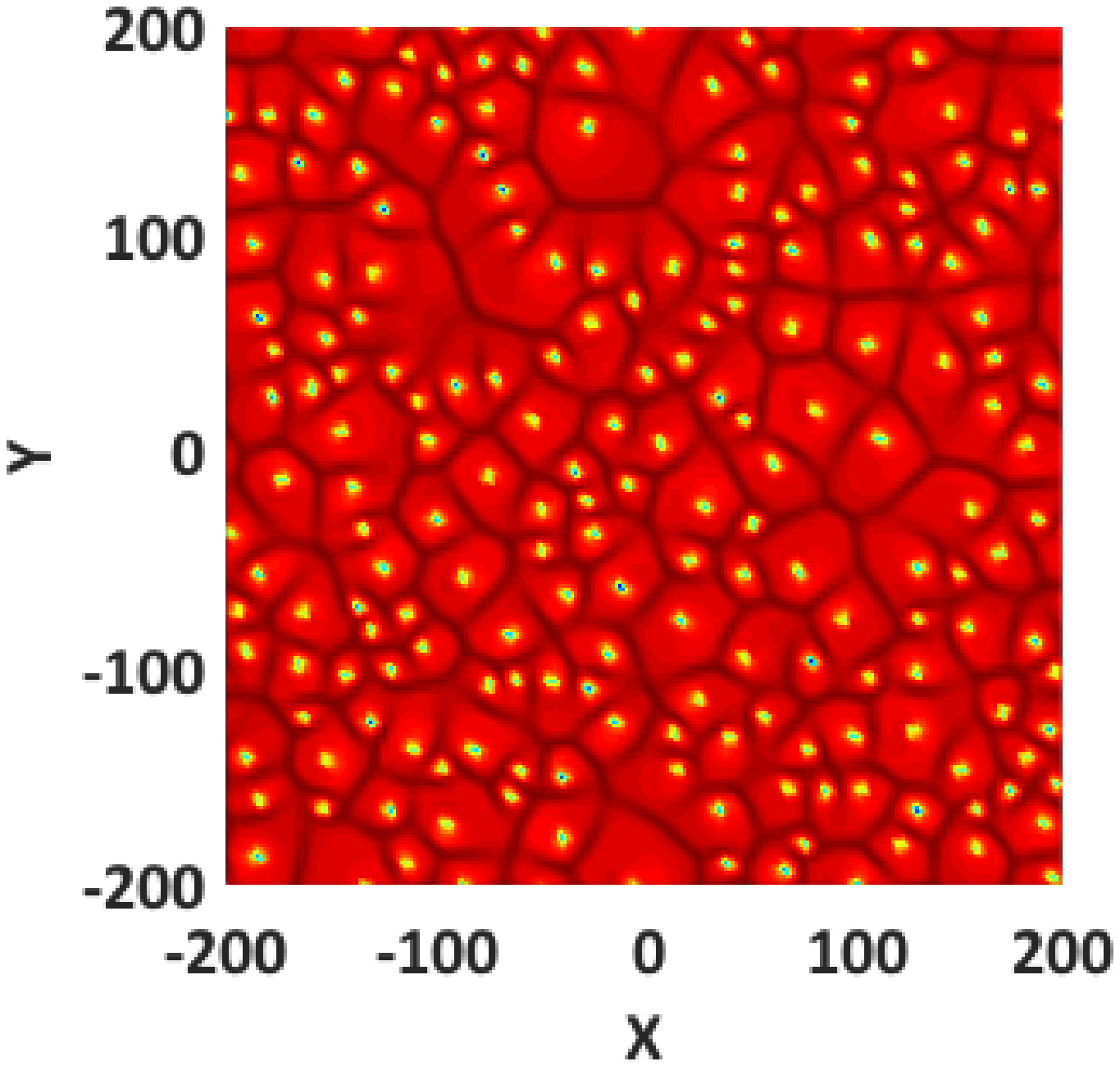}}\\
       \subfloat[] {\includegraphics[width=1.8 in]{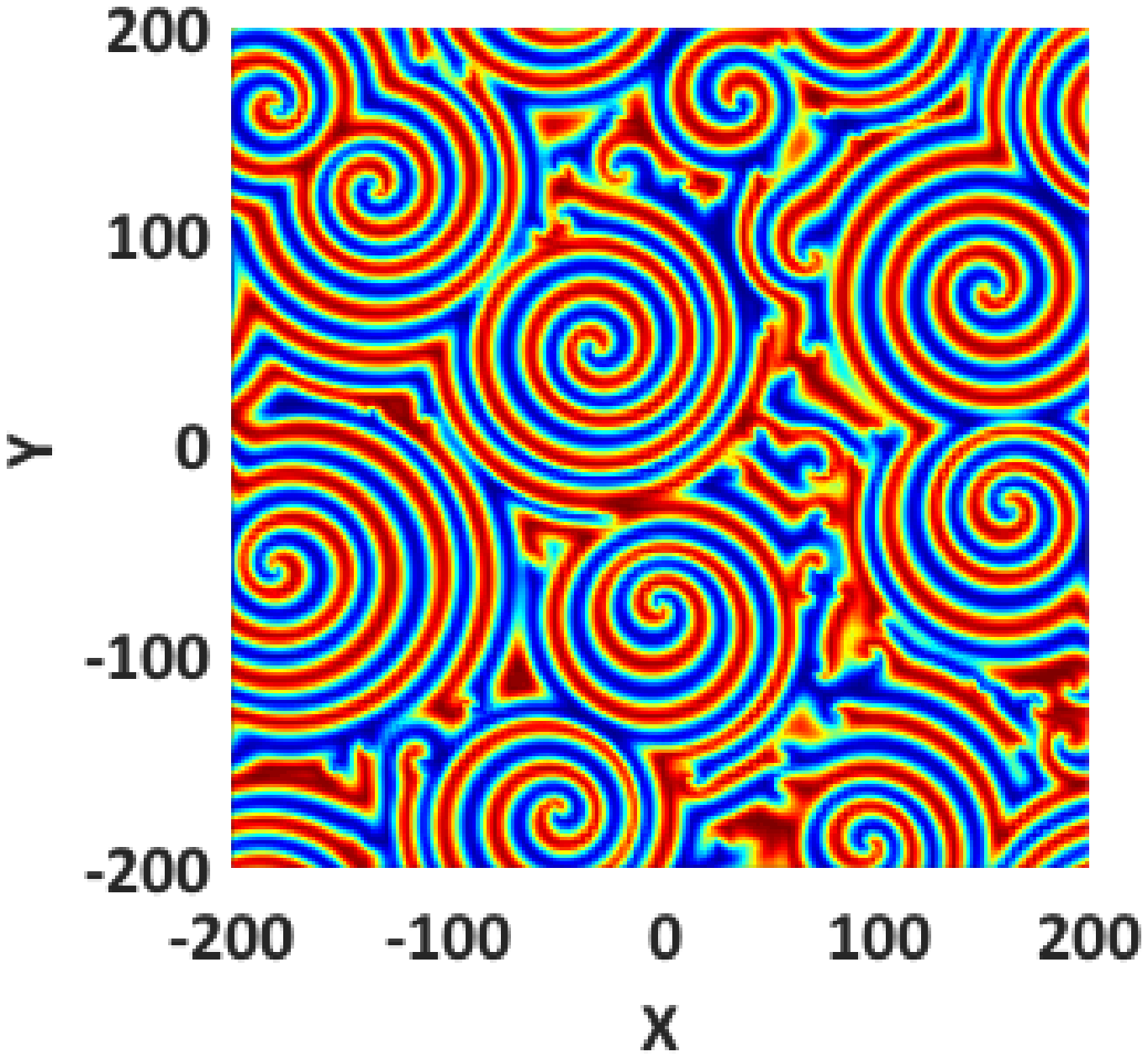}}
       \subfloat[] {\includegraphics[width=1.8 in]{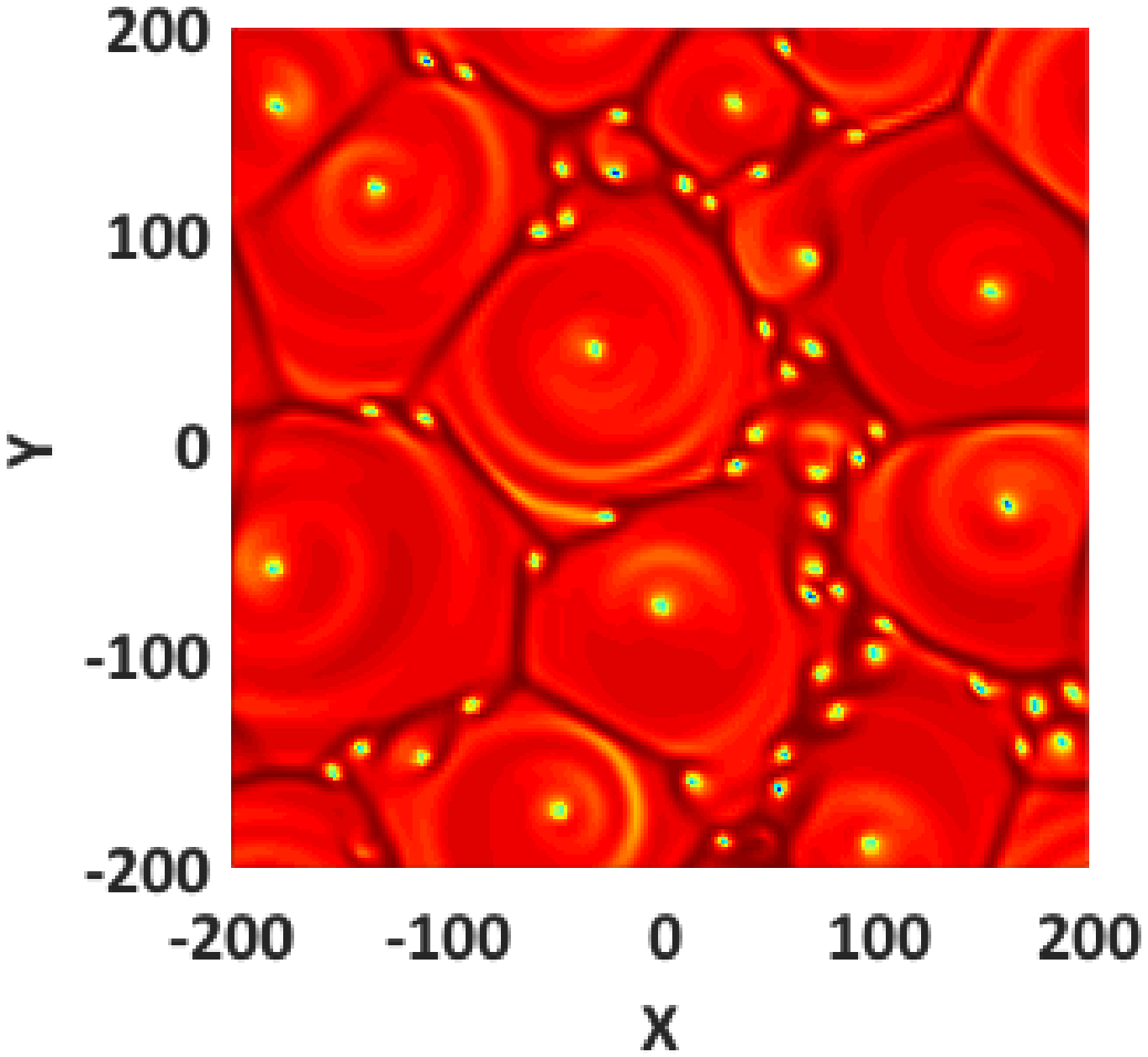}}

      \subfloat[] {\includegraphics[width=1.8 in]{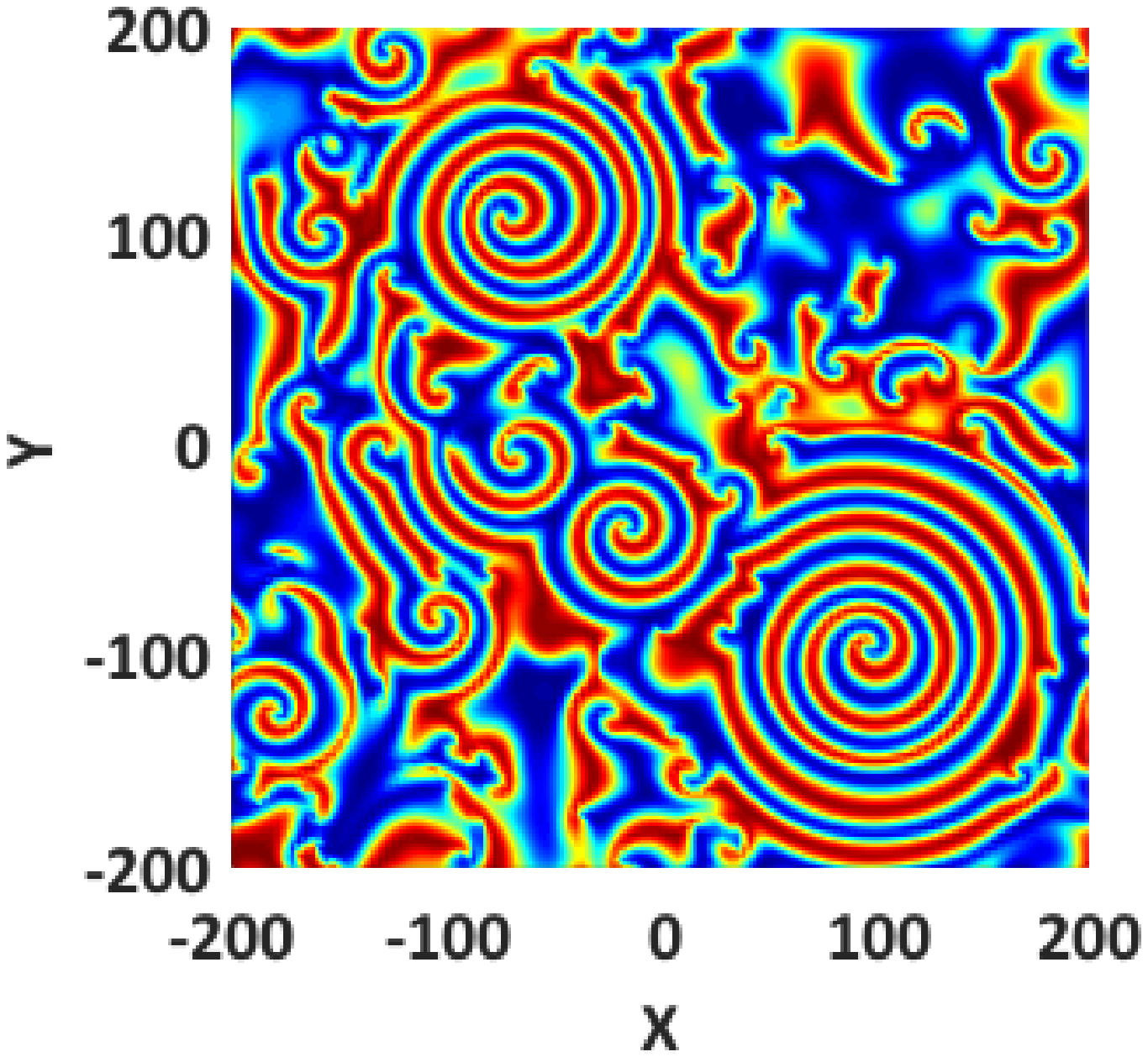}}
      \subfloat[] {\includegraphics[width=1.8 in]{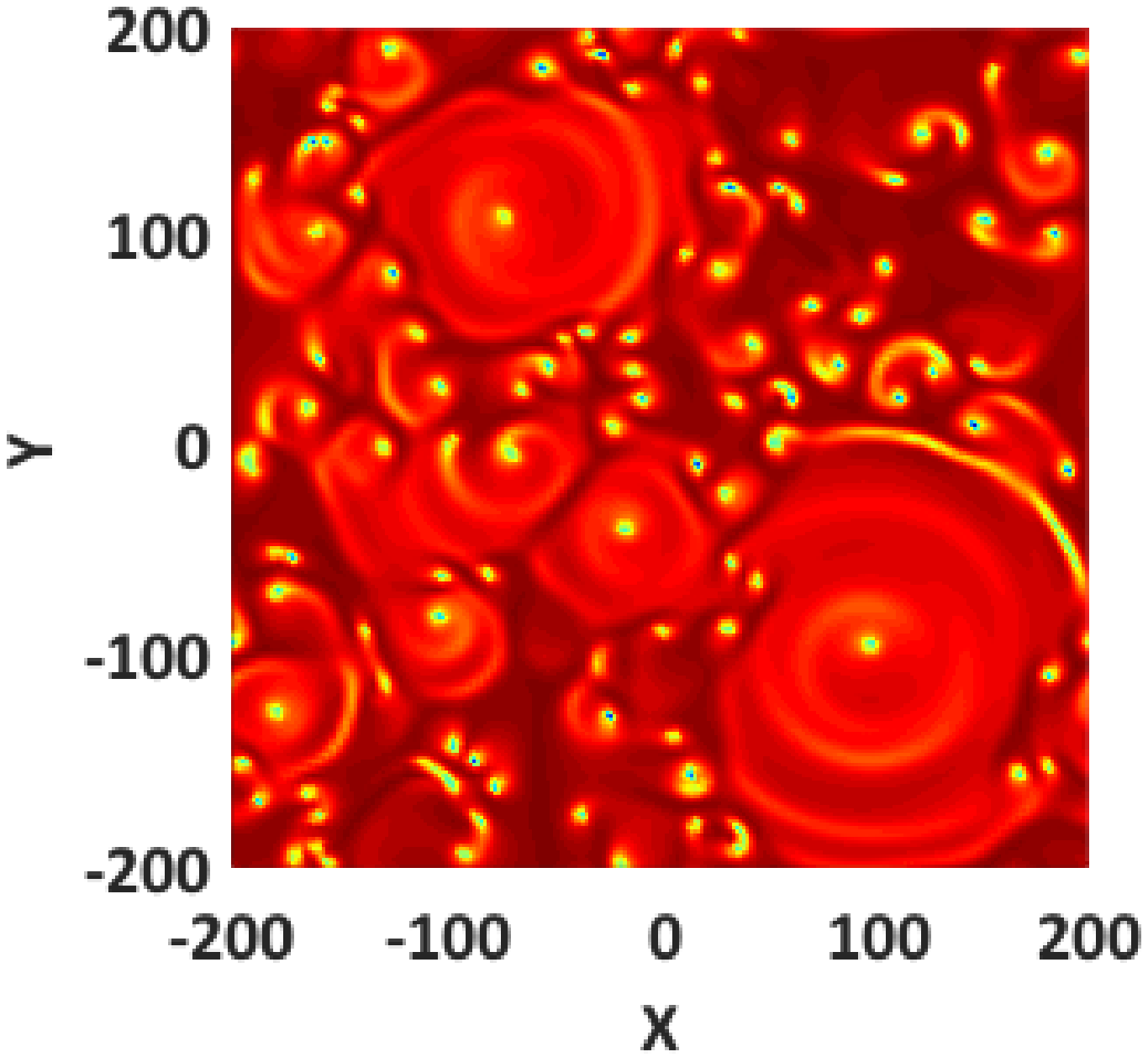}}\\
      \subfloat[] {\includegraphics[width=1.8 in]{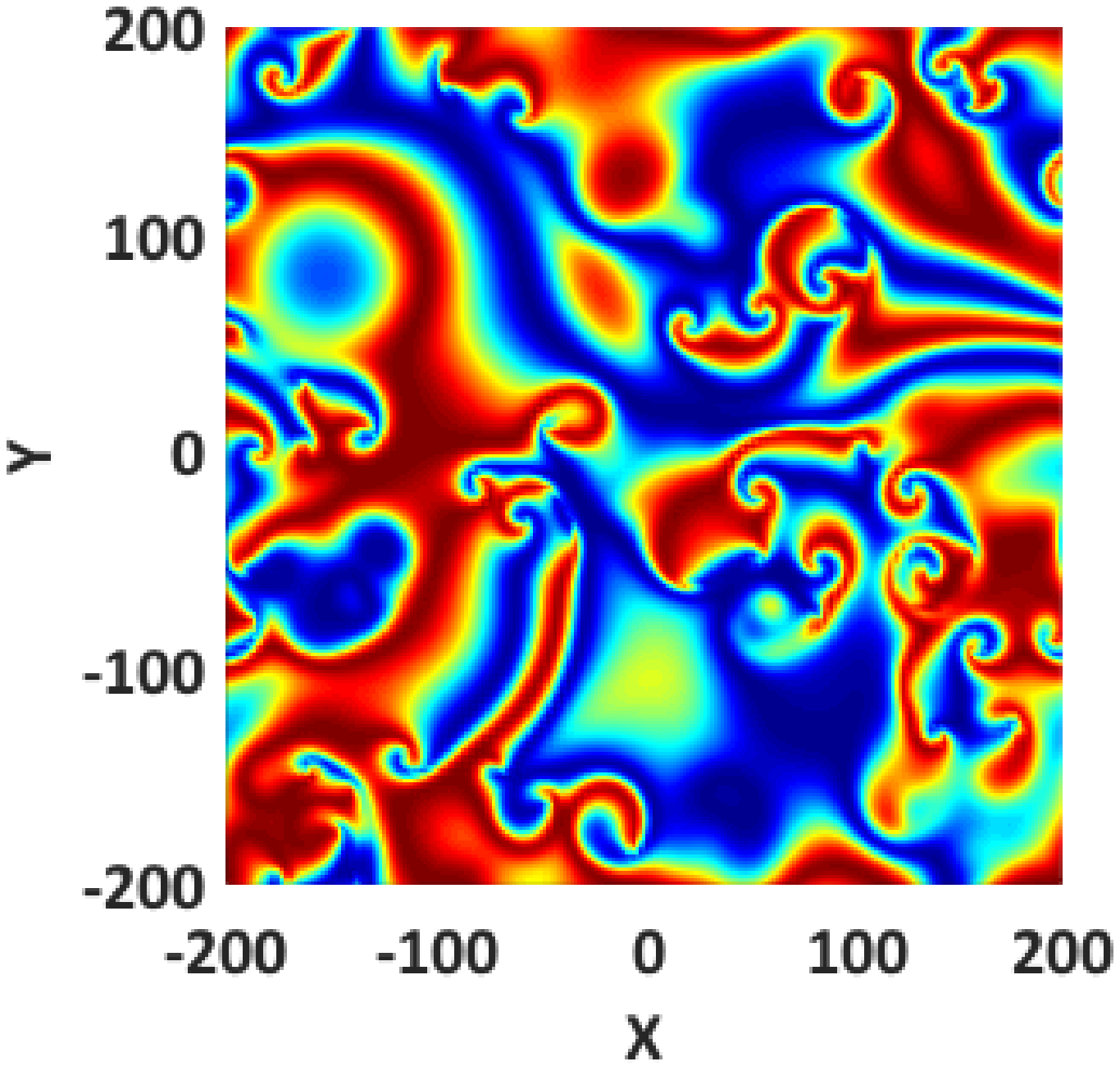}}
      \subfloat[] {\includegraphics[width=1.8 in]{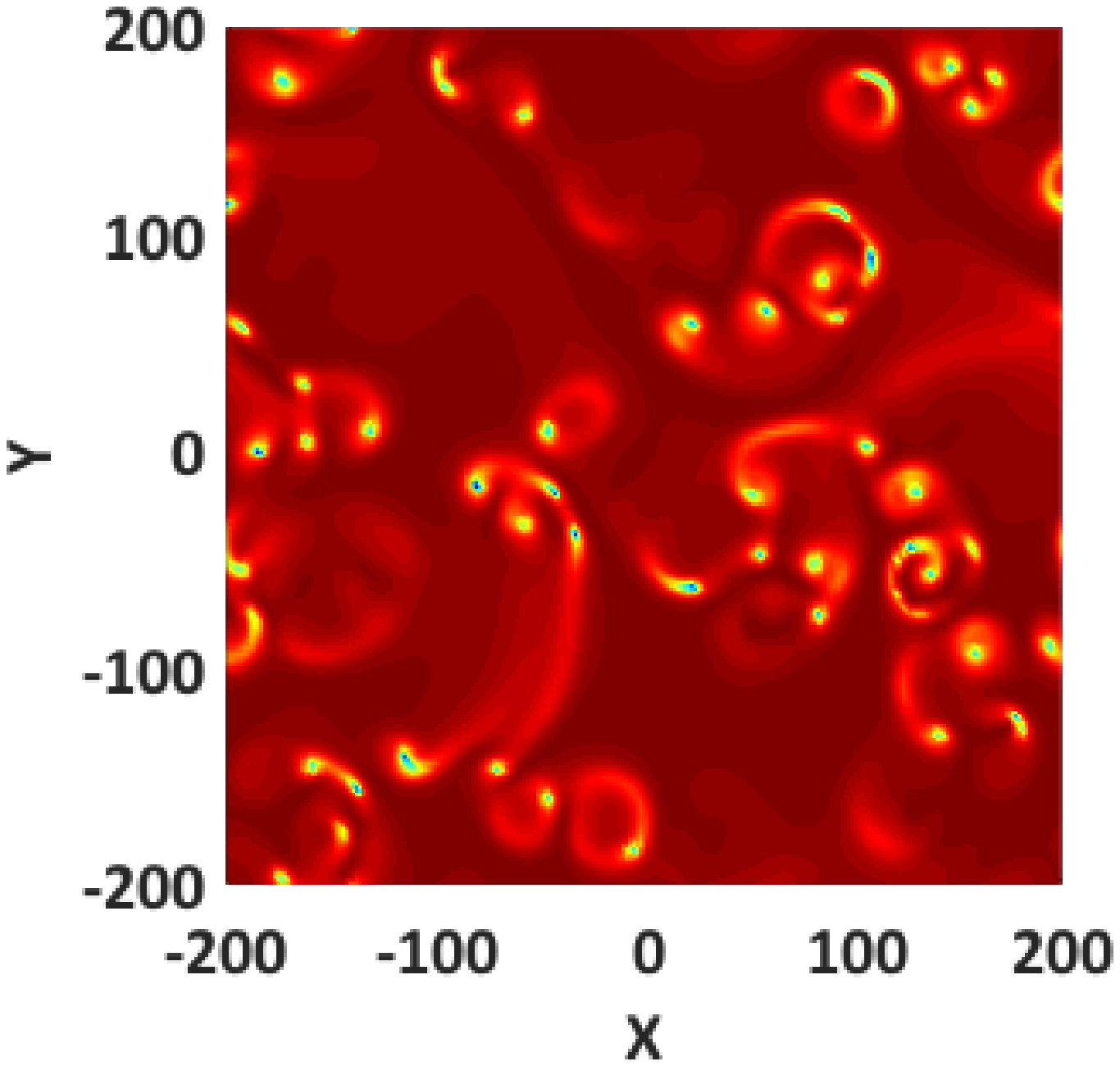}}
\caption{Snapshots of solutions of \eqref{CGLERS} for the parameters $c_1=0.55$,
    $c_2=-0.45$ and the system size $L=400$. Left panel, $Re(W)$, and right panel, $|W|$,
    with $\alpha=2$ for (a) and (b); $\alpha=1.46$ for (c) and (d); $\alpha=1.4$ for (e) and (f) and $\alpha=1.3$ for (g) and (h).}\label{spiral}
\end{center}
\end{figure}

We start with the values of $c_1$ and $c_2$ for which in the case of
normal diffusion, CGLE leads to the frozen states (for instance see
Fig. \ref{spiral}.a, \ref{spiral}.b) with cellular structures (Fig.
\ref{spiral}.b). It is seen in Fig. \ref{spiral}.a, that there are
no apparent spiral waves in the case of normal diffusion for the
selected parameters. With the same parameters in the case of
superdiffusion, as $\alpha$ decreases, cellular structures with
larger size are formed and then with further decrease of $\alpha$,
spiral waves emerge (Fig. \ref{spiral}.c). Fig. \ref{spiral}.f shows
that as $\alpha$ becomes smaller, the domain walls (shock lines)
almost melt and as can be seen in Fig. \ref{spiral}.e a mixed state
appears in which some spirals live in a disordered sea. Finally the
spiral patterns as well as the cellular structures disappear
completely as one further reduces the value of $\alpha$. In this
stage we obtain spatio-temporal patterns for which $|W|$ is not
stationary in time and becomes zero at some points (Fig.
\ref{spiral}.h). This is the fully developed defect turbulence
regime according to its definition. Therefore, numerical simulations
show that superdiffusion can create and annihilate spiral patterns
and can make a transition from one regime to another. In fact the
magnitude of diffusion and dispersion terms become larger in the
presence of superdiffusion and the competition between the three
terms in r.h.s of \eqref{CGLERS} changes in comparison with the case
of normal diffusion and can even lead to the probable change of
regimes.

\begin{figure}[H]
    \begin{center}
        \subfloat[]{\includegraphics[width=1.8 in]{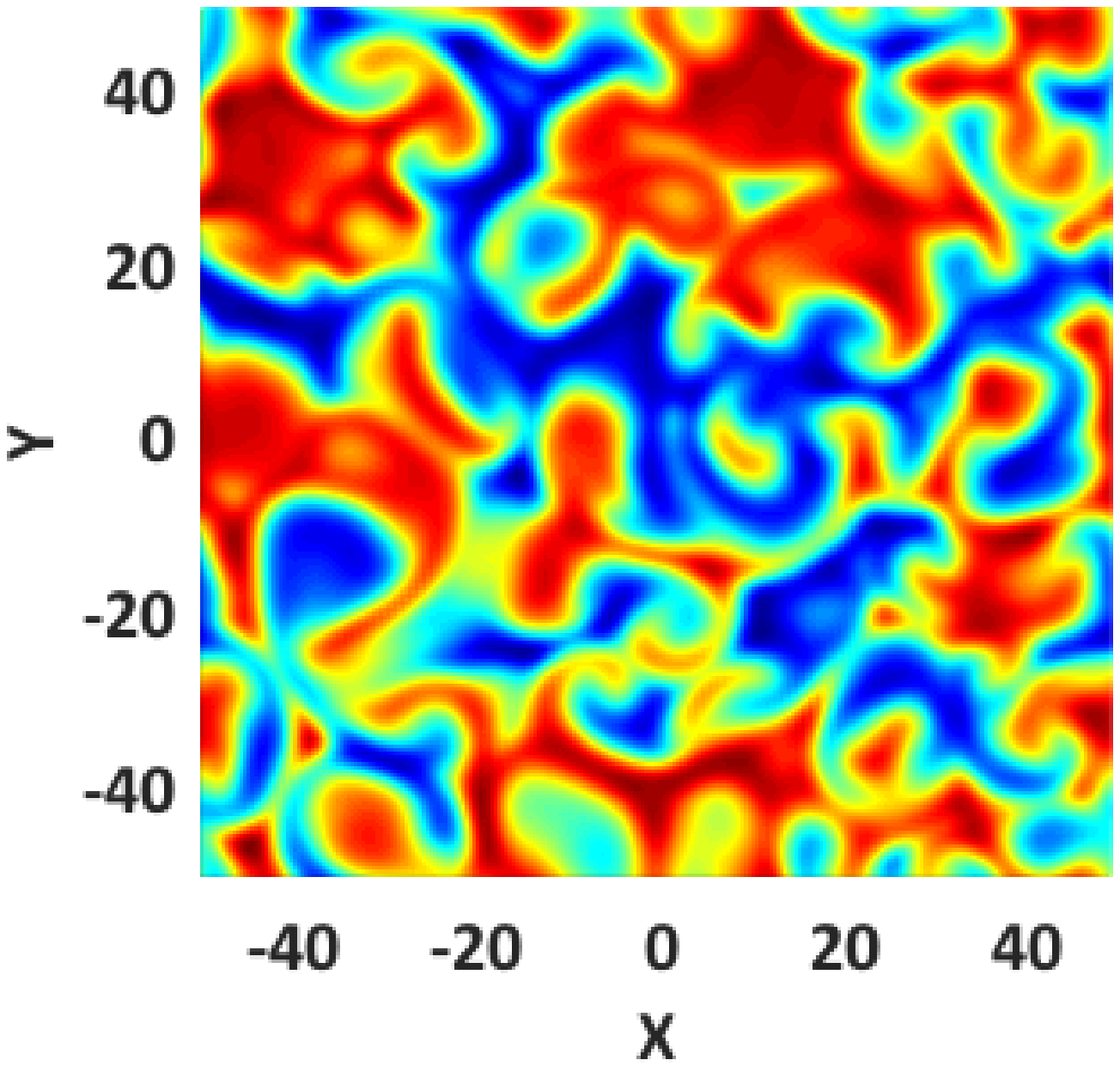}}
        \subfloat[]{\includegraphics[width=1.8 in]{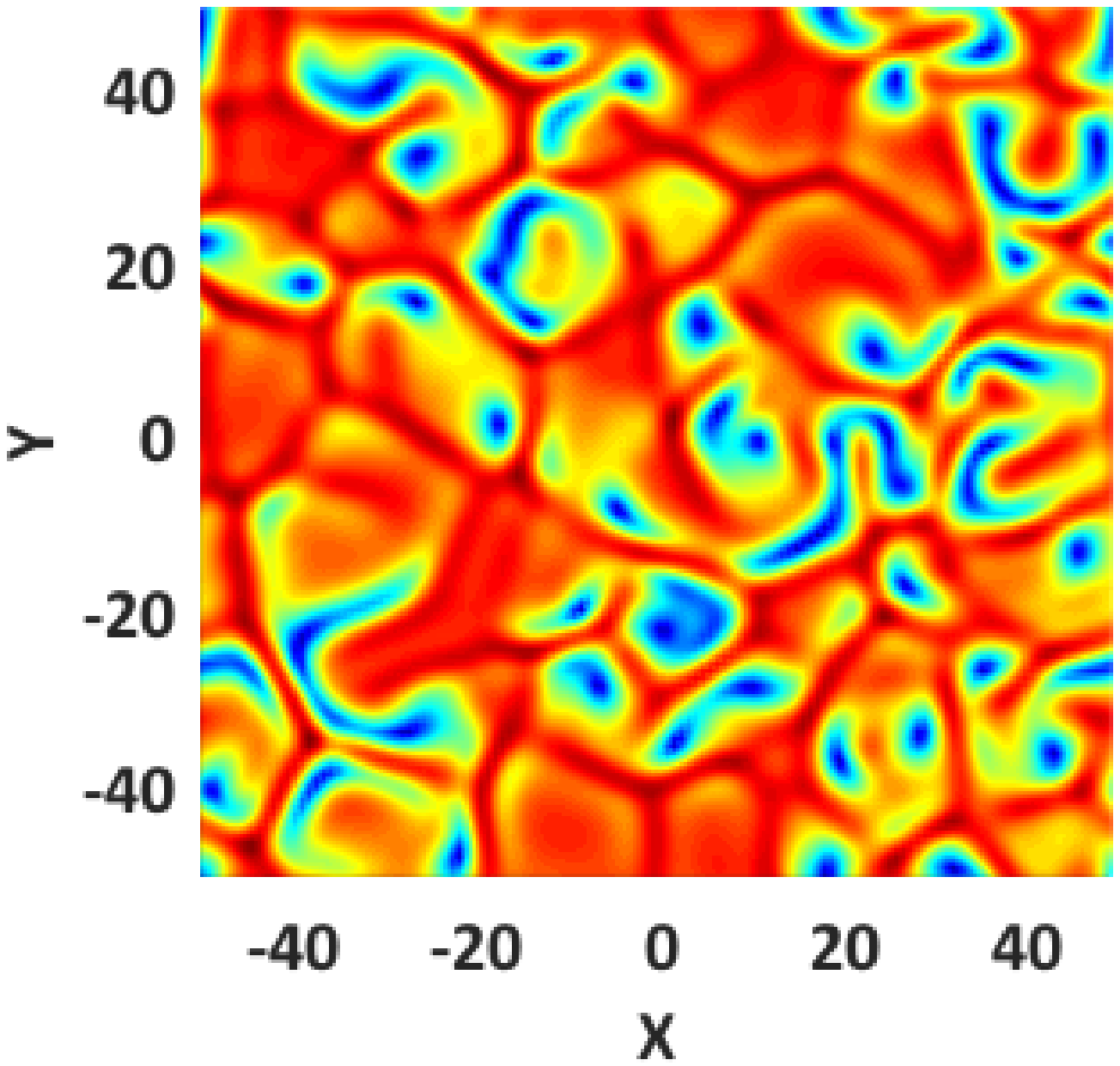}}\\
        \subfloat[]{\includegraphics[width=1.8 in]{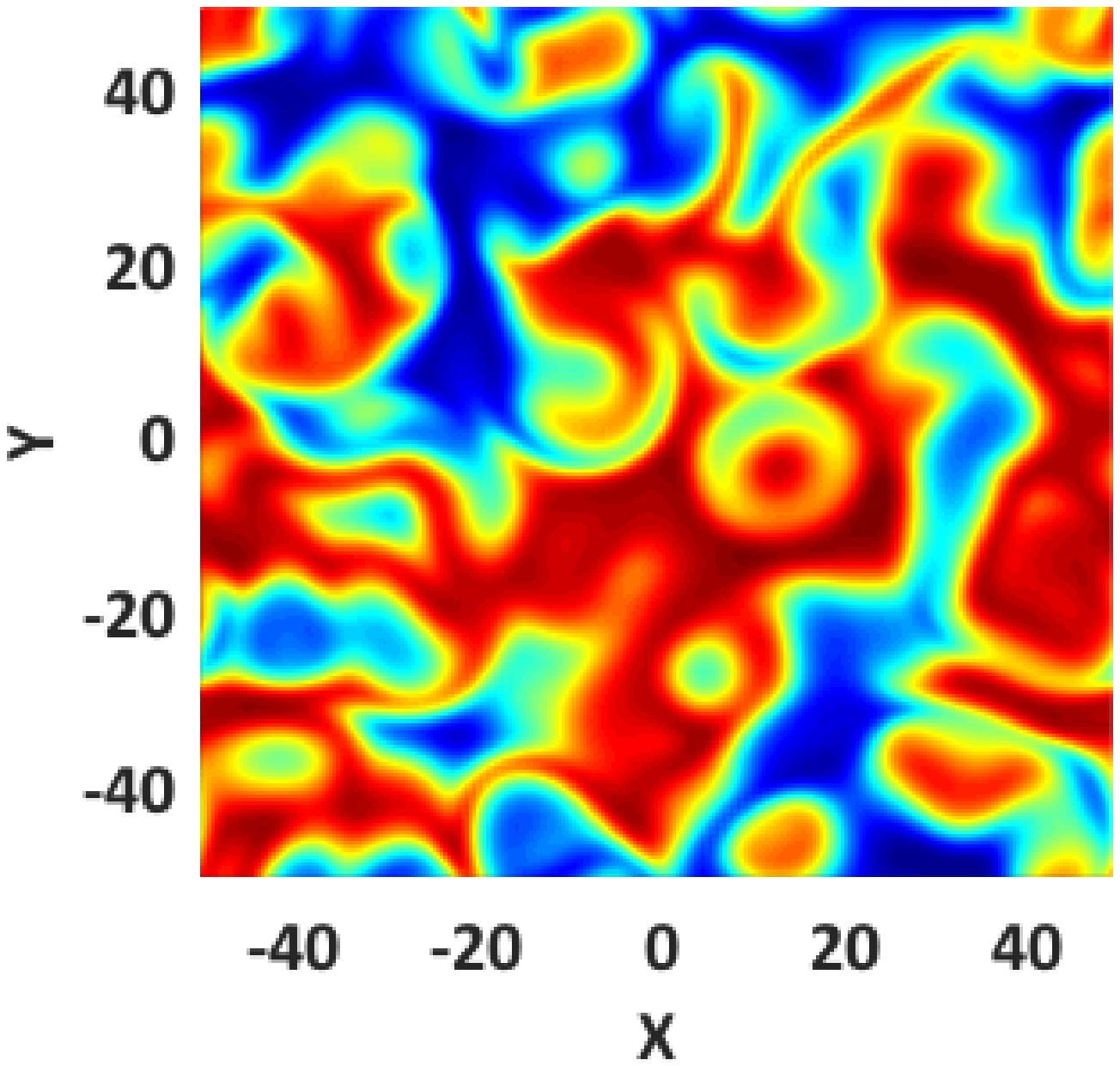}}
        \subfloat[]{\includegraphics[width=1.8 in]{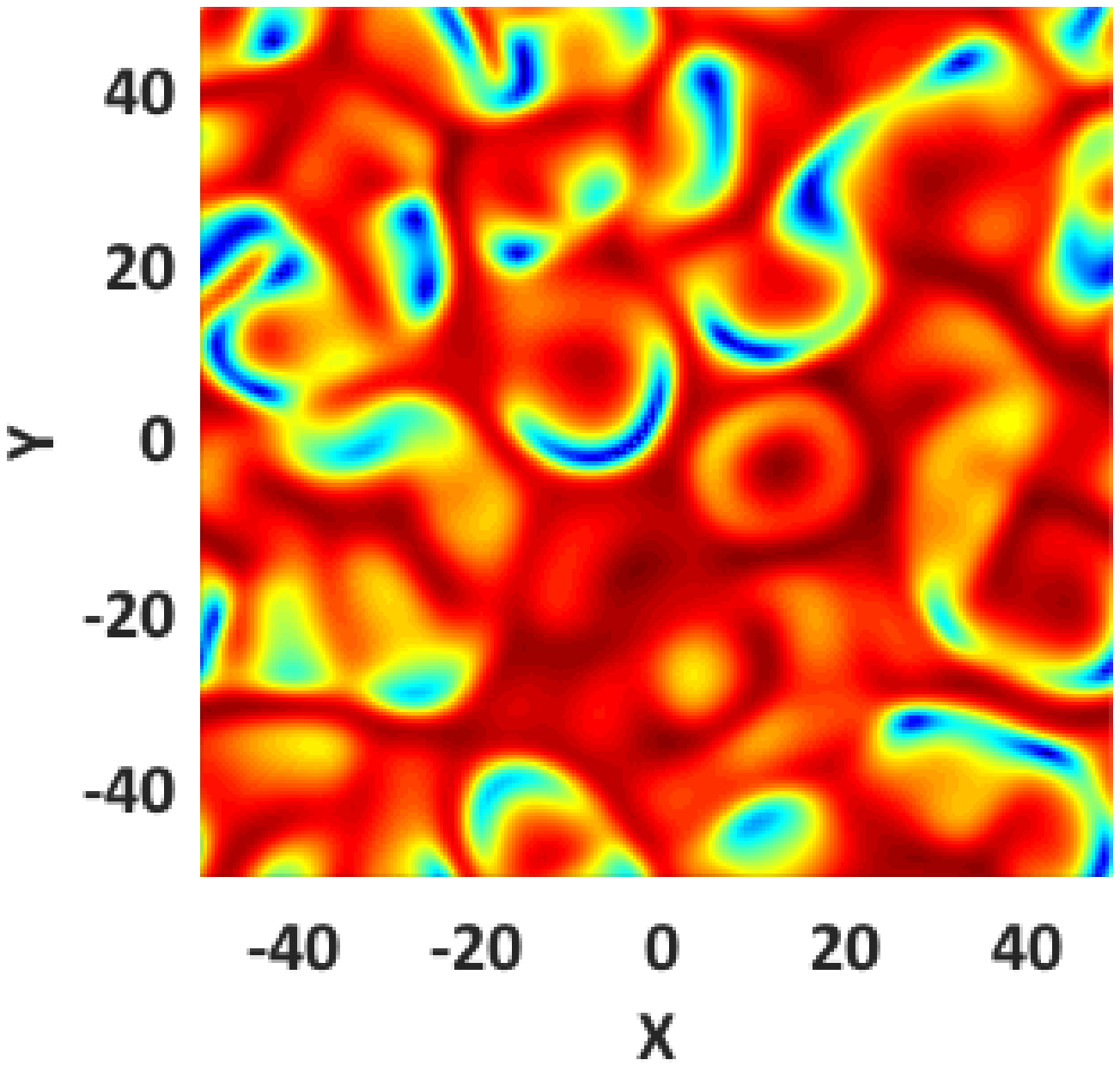}}
        \caption{ Defect turbulence regime: Snapshots of solutions of \eqref{CGLERS} for the parameters $c_1=1$,
            $c_2=-1.2$ and the system size $L=100$. Left panel, $Re(W)$, and right
            panel, $|W|$, with  $\alpha=2$ for (a) and (b) and $\alpha=1.1$ for (c) and (d).  }\label{defect}
    \end{center}
\end{figure}

Numerical simulations indicate that in the defect turbulence regime,
density of defects reduces gradually by the decrease of fractional
order, $\alpha$. For instance in Fig. \ref{defect} the parameters
are chosen in such way that we are in the defect turbulence regime
of CGLE with a large density of defects (Fig. \ref{defect}.a,
\ref{defect}.b) and  it is seen that the density of defects
decreases in the presence of superdiffusion  (Fig. \ref{defect}.c,
\ref{defect}.d).

\begin{figure}[H]
    \begin{center}
    \subfloat[]{\includegraphics[width=1.8 in]{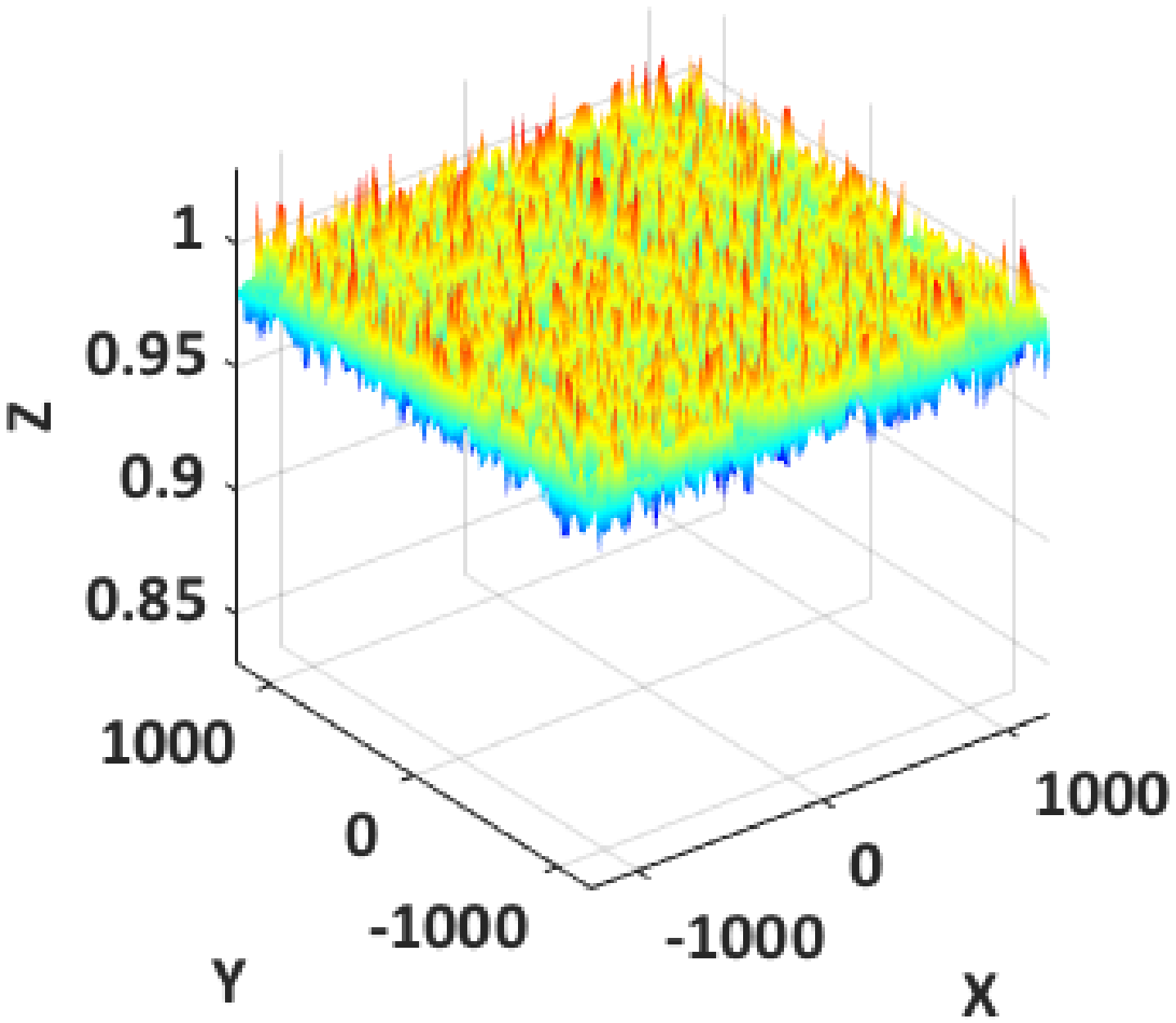}}
    \subfloat[]{\includegraphics[width=1.8 in]{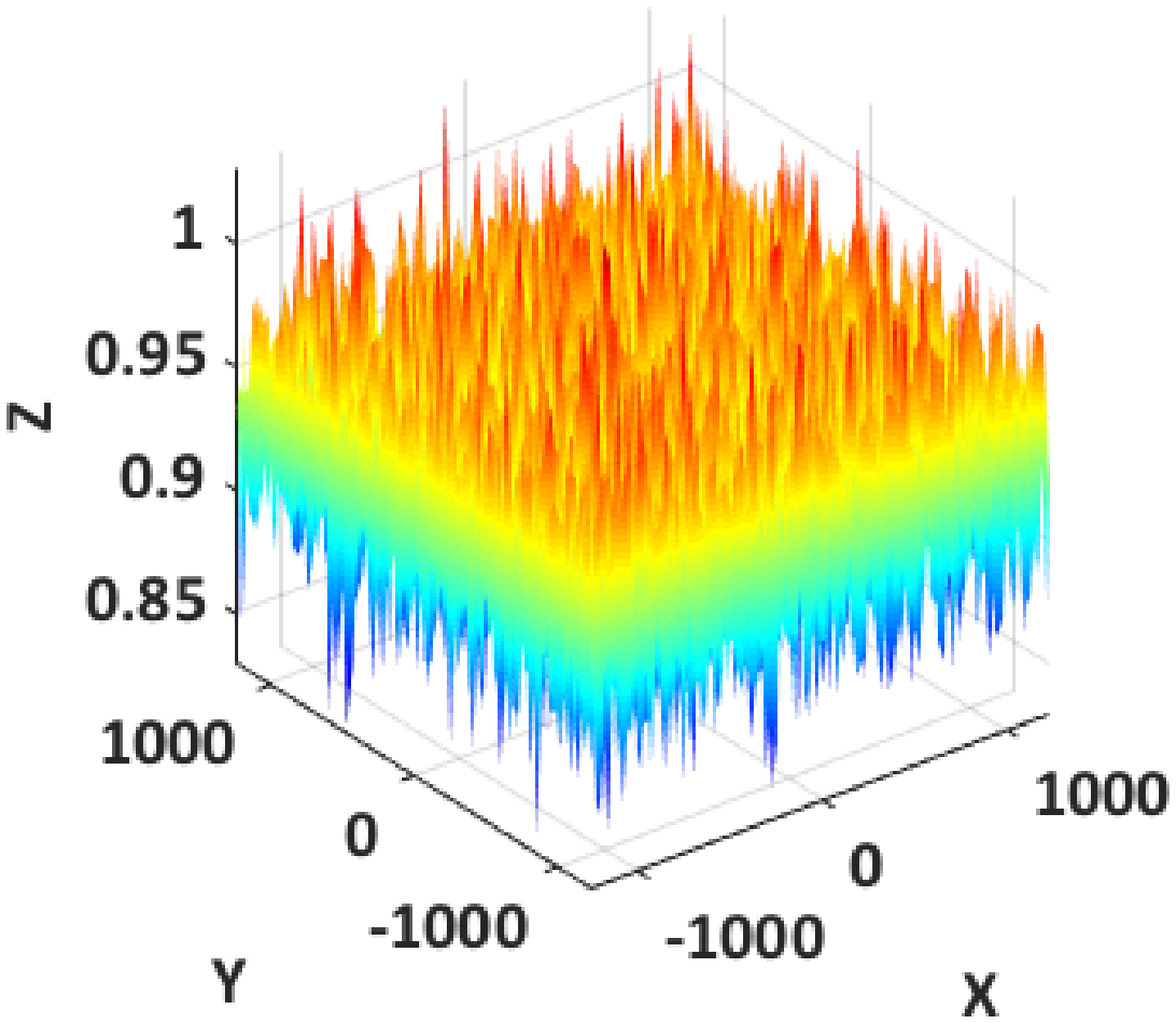}}
        \caption{ Phase turbulence regime: Snapshots of solutions of \eqref{CGLERS}  for the parameters $c_1=2$,
            $c_2=-1$ and the system size $L=2500$. Plot of $|W|$ for (a) $\alpha=2$ and (b) $\alpha=1.1$. }\label{phase}
    \end{center}
\end{figure}

For the phase turbulence regime, as can be seen in Fig. \ref{phase}
as an example, the amplitude of fluctuations increases in the
superdiffusion case.

\begin{figure}[H]
    \begin{center}
        \subfloat[]{\includegraphics[width=1.8 in]{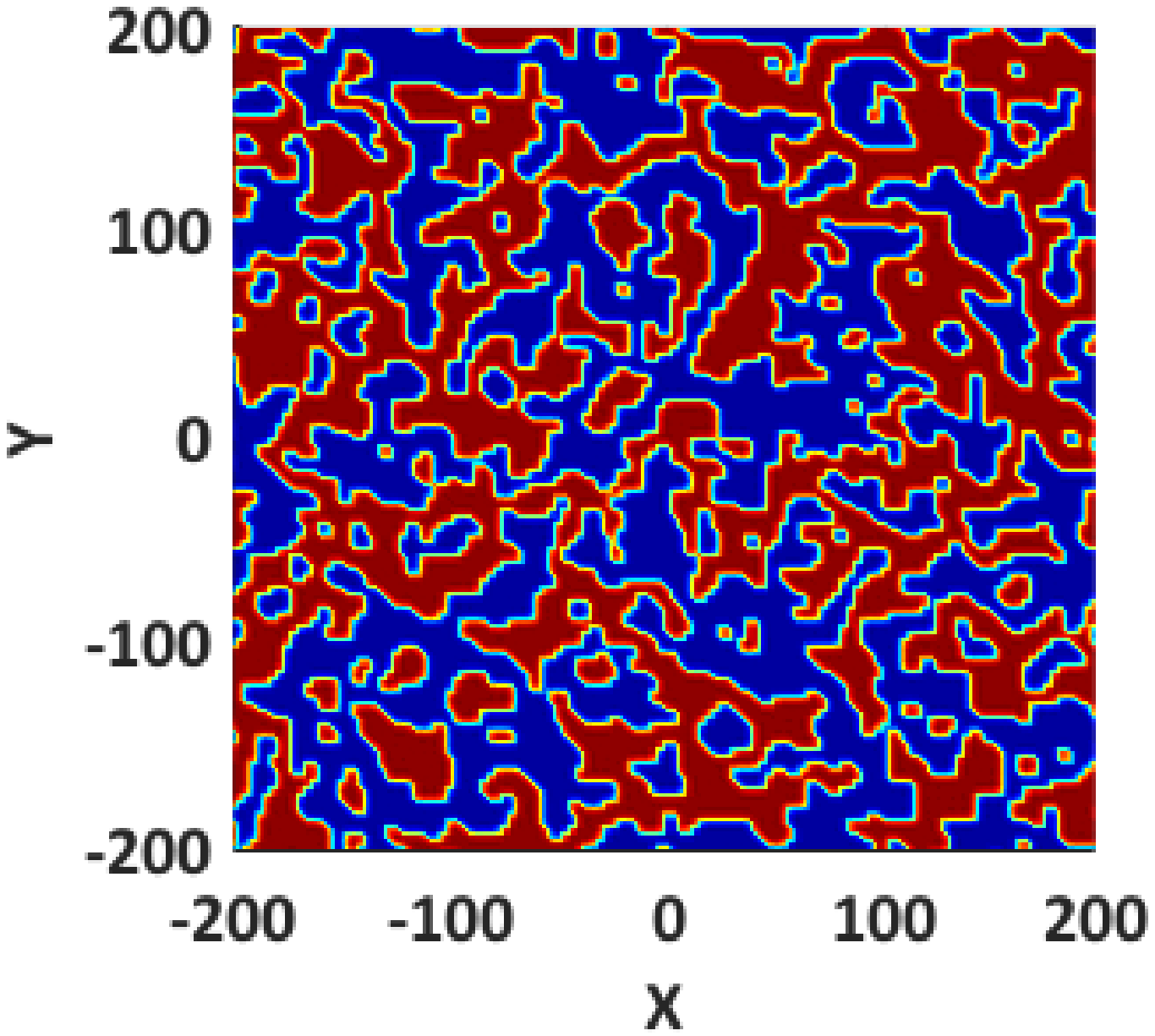}}
        \subfloat[]{\includegraphics[width=1.8 in]{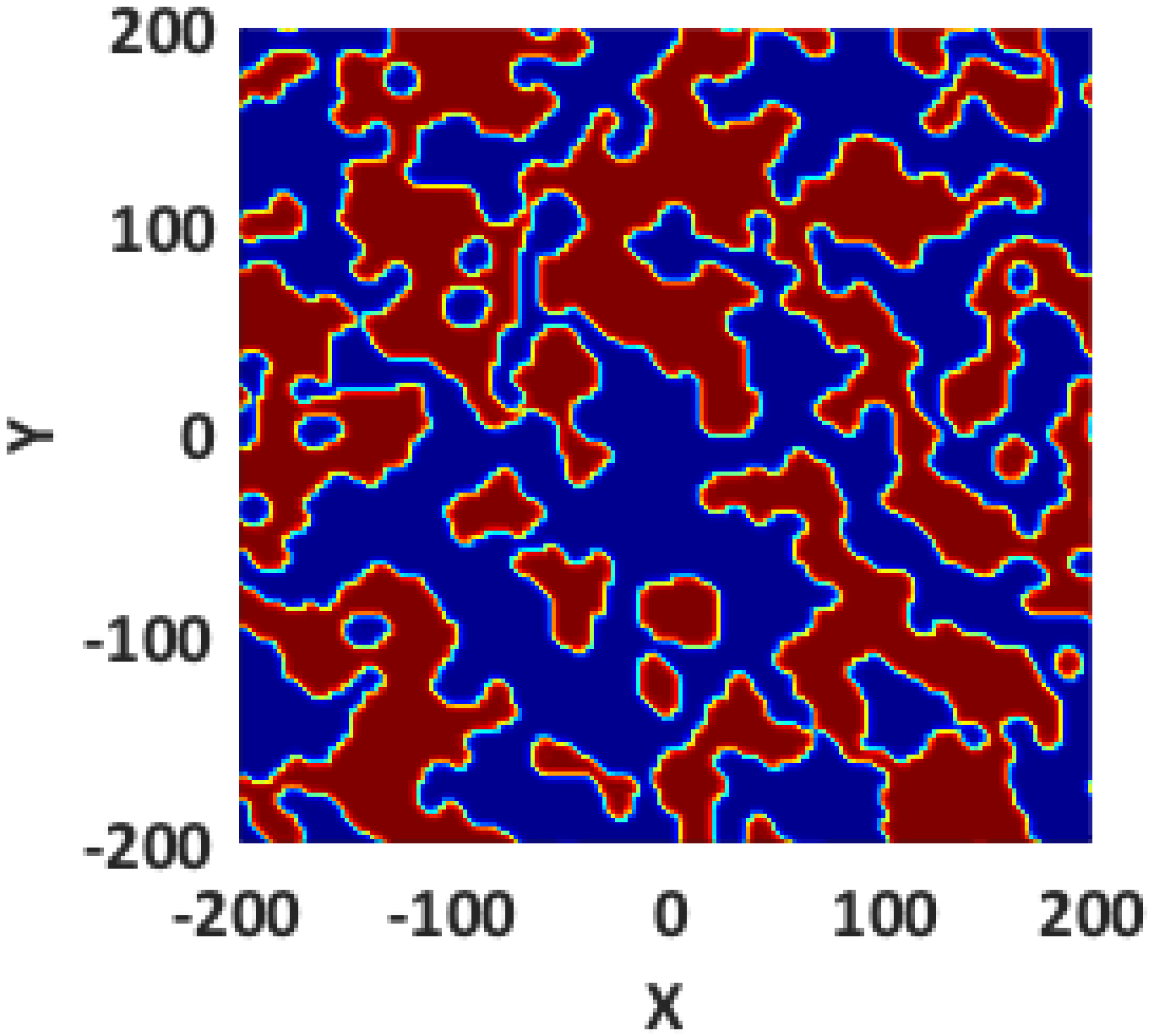}}
        \subfloat[]{\includegraphics[width=1.8 in]{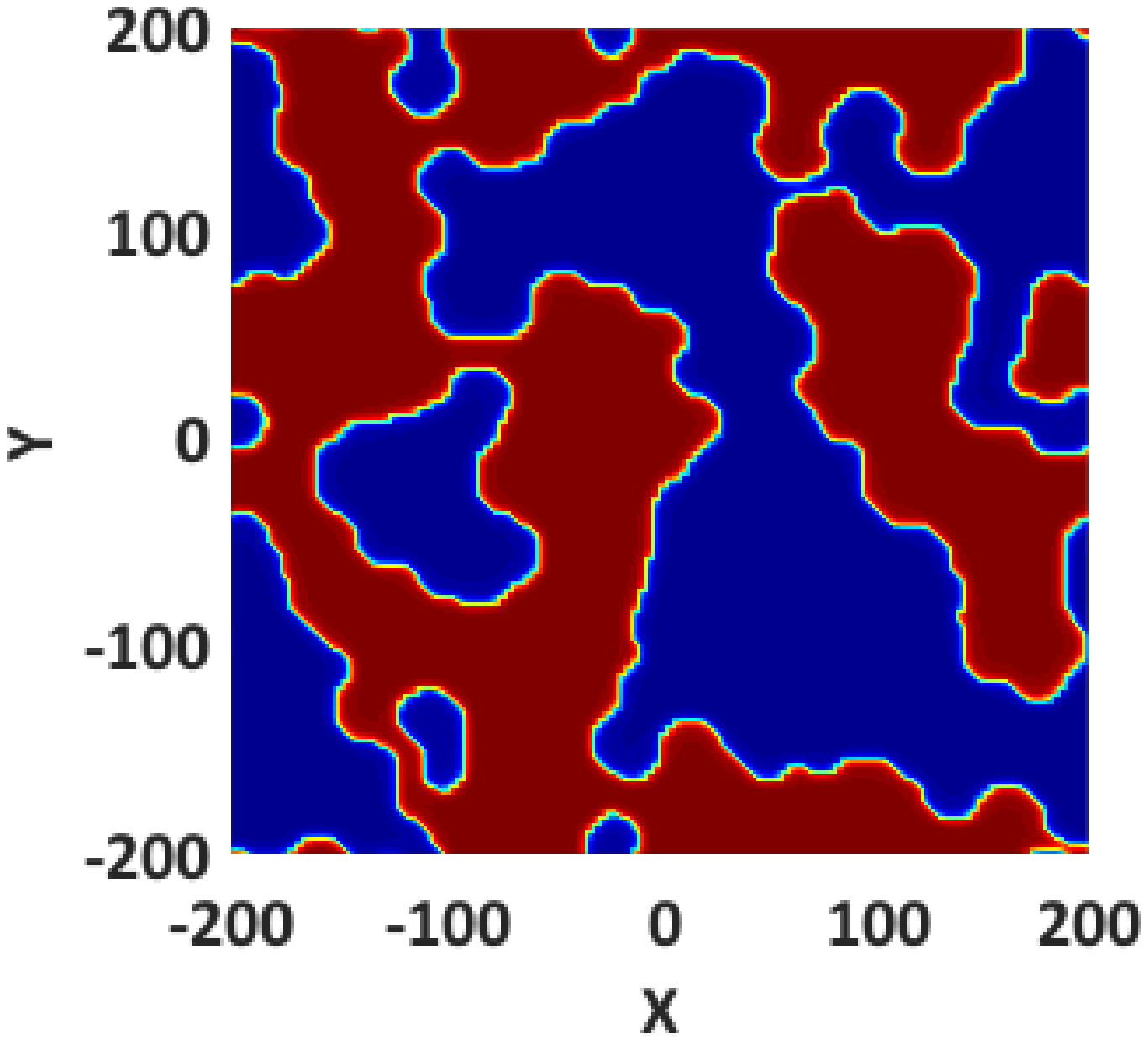}}\\
        \subfloat[]{\includegraphics[width=1.8 in]{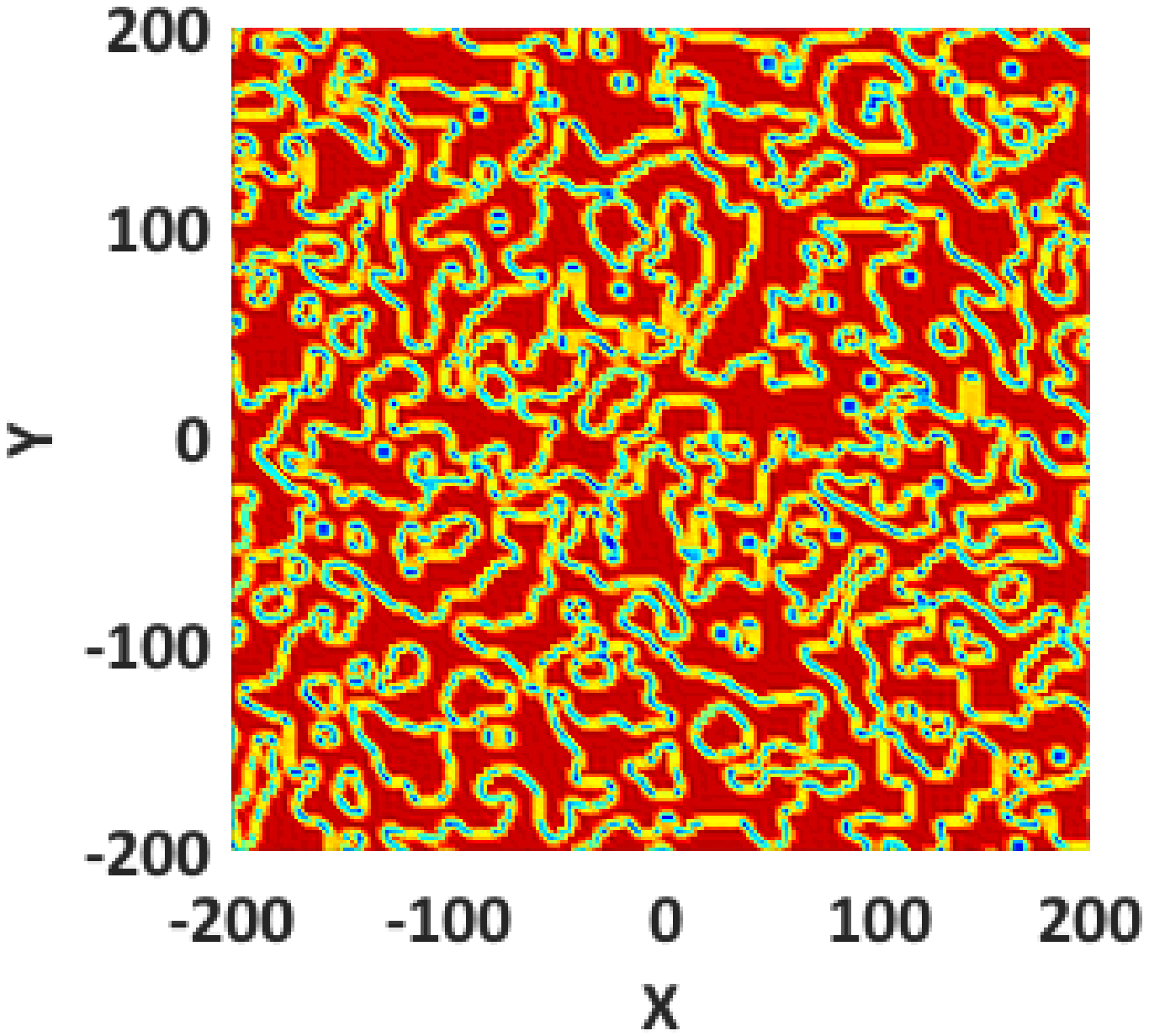}}
        \subfloat[]{\includegraphics[width=1.8 in]{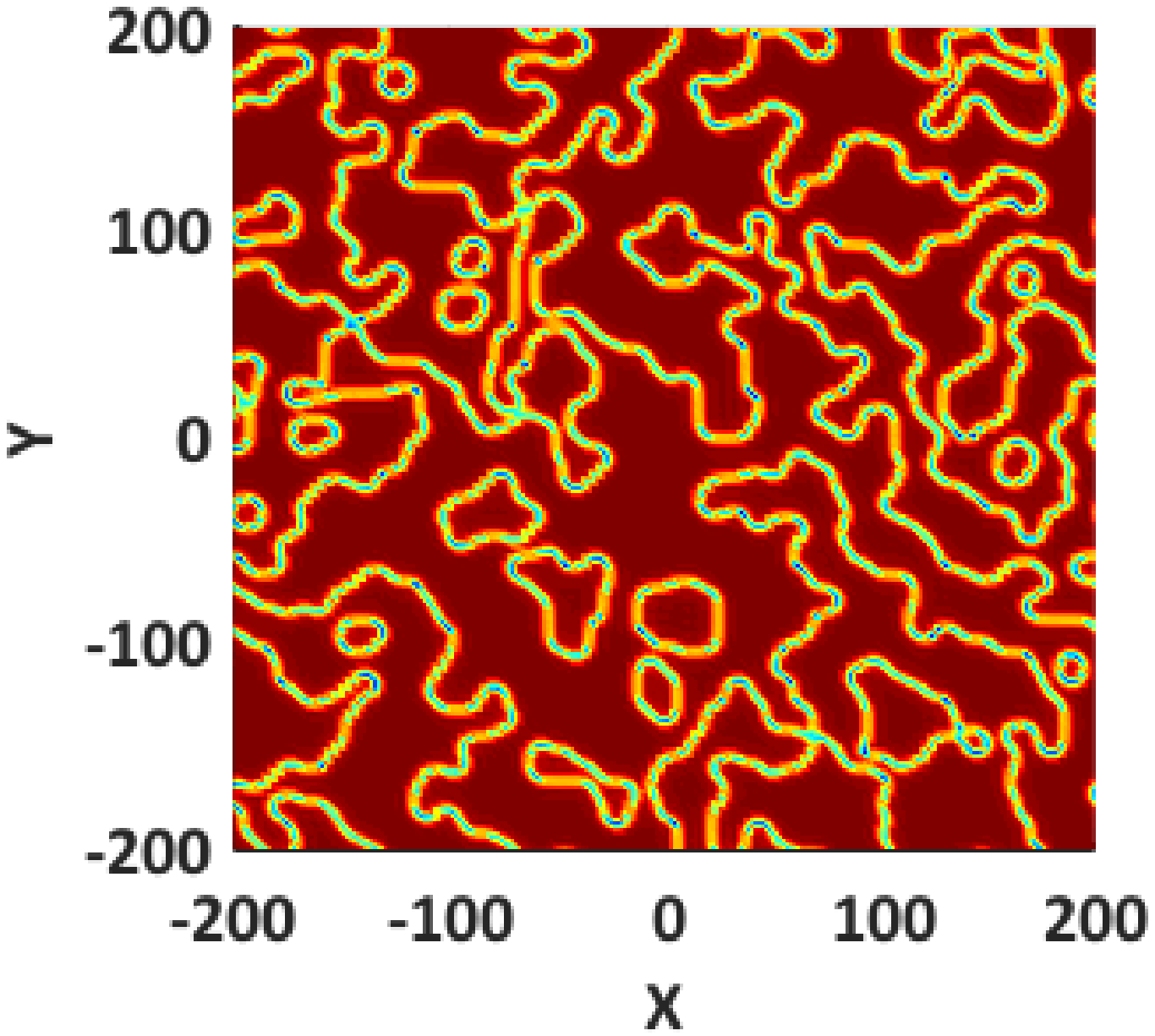}}
        \subfloat[]{\includegraphics[width=1.8 in]{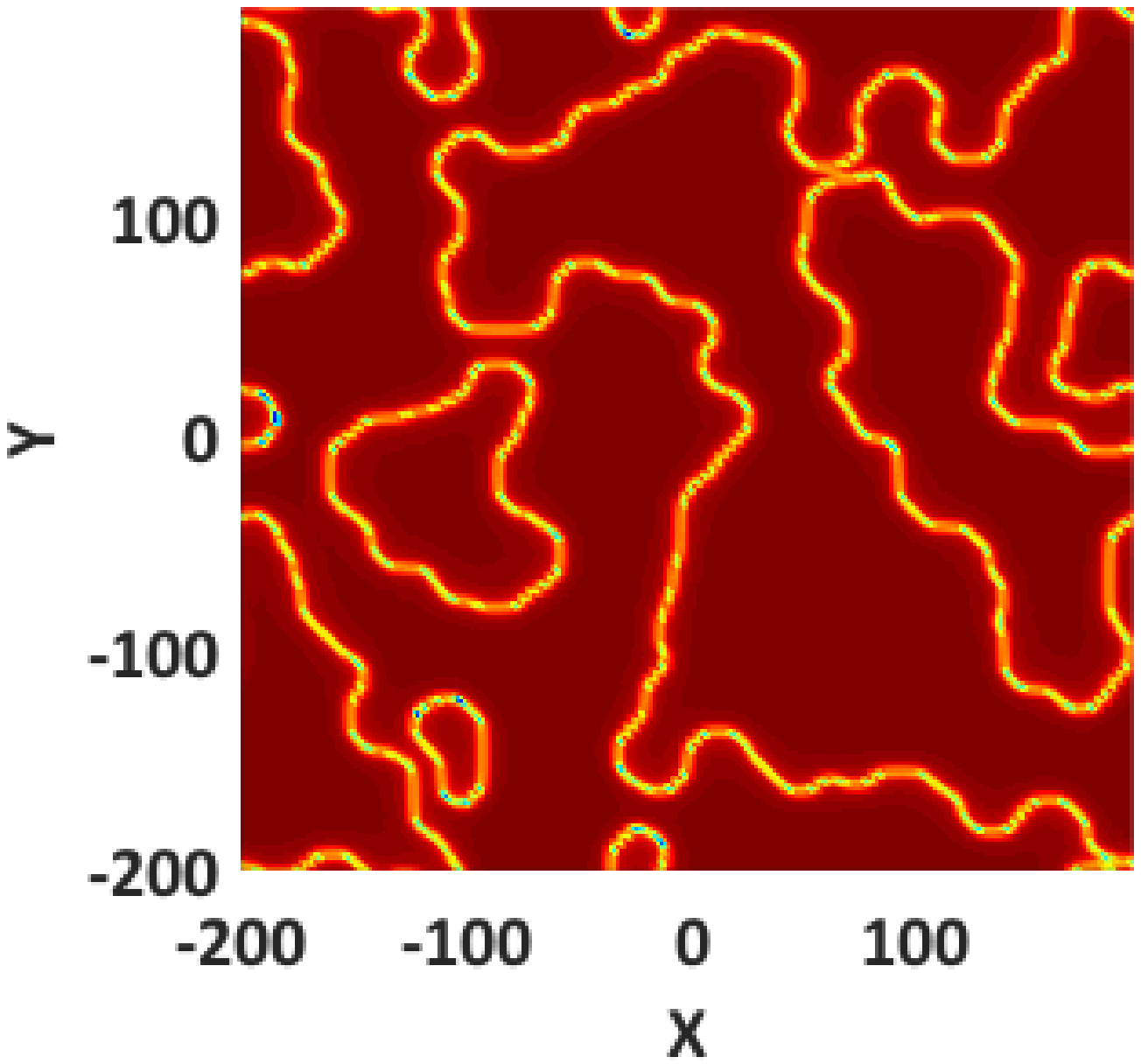}}
        \caption{Snapshots of solutions of \eqref{CGLERS} for the parameters $c_1=c_2=0$ and the system size $L=400$.
        Upper panel, $Re(W)$,  and lower panel, $|W|$, with $\alpha=2$ for (a) and (d); $\alpha=1.5$ for (b) and (e) and $\alpha=1.1$ for (c) and (f).  }\label{RGLE}
    \end{center}
\end{figure}

After investigation of different regimes in FCGLE, we proceed by
studying the interesting limit of this equation, $c_1, c_2
\rightarrow 0$, which is the time-dependent {\it fractional} real
Ginzburg-Landau equation. Dynamics of this equation results in
stationary patterns for both $Re(W)$ and $|W|$ (Fig. \ref{RGLE}).
For the case of normal diffusion, $\alpha=2$, FRGLE reduces to real
Ginzburg-Landau equation (RGLE) which has been used to describe the
spinodal decomposition in critical quench \cite{Petschek}.
Simulations of time-dependent RGLE results in domain patterns that
can be seen in figures \ref{RGLE}.a and \ref{RGLE}.d. For the
superdiffusion case, i.e. $\alpha\neq 2$, the domain patterns also
form for the FRGLE but, as is clear from figures \ref{RGLE}.b,
\ref{RGLE}.e, \ref{RGLE}.c and \ref{RGLE}.f, larger size domains
appear as $\alpha$ decreases.

\section{Turing instability in reaction-superdiffusion systems}

In the previous section we studied the behavior of a
reaction-superdiffusion system near a Hopf instability. In this
section we are going to describe a two-component
reaction-superdiffusion system in the vicinity of a Turing
instability. In fact, a generalized thermodynamic potential, free
energy, will be eventually presented which governs the stability,
the dynamics and the fluctuations of reaction-superdiffusion systems
near the Turing bifurcation. To reach this goal we include a local
noise term in the linearized system of reaction-superdiffusion
equations. Note that the reason for considering a two-component
system in this section is that we want to present an analytic
expression for the free energy.

Recall the general equation \eqref{Reacdif} for a n-component
reaction-superdiffusion system. For a two-component system ${\bf
X}({\bf r},t)$ and ${\bf F}({\bf X};\mu)$ are given by ${\bf X}({\bf
r},t)=\left( \begin{array}{cc} X_1({\bf r},t)\\ X_2({\bf r},t)
\end{array} \right)$ and ${\bf F}({\bf X};\mu)=\left(
\begin{array}{cc} F_1(X_1,X_2;\mu)\\ F_2(X_1,X_2;\mu) \end{array}
\right)$ and therefore \eqref{Reacdif} in the presence of noise is
written as
\begin{equation} \label{activ-inhabit}
\left\{
\begin{array}{cc}
\frac{\partial X_1}{\partial t}=D_{\alpha1} \nabla^\alpha
X_1+F_1(X_1,X_2;\mu)+\xi_1({\bf r},t)\\
\frac{\partial X_2}{\partial t}=D_{\alpha2} \nabla^\alpha
X_2+f_2(X_1,X_2;\mu)+\xi_2({\bf r},t)
\end{array}
\right.,
\end{equation}
where $\xi_1$ and $\xi_2$ are noise terms that generally depend on
$\bf r$ and $t$. This  two-component reaction-superdiffusion system
is known as activator-inhibitor system where $X_1$ and $X_2$ are
activator and inhibitor, respectively. We consider the noise terms
to be Gaussian white noises with the following properties
\[
\langle \xi_1({\bf r},t) \rangle=\langle \xi_2({\bf r},t) \rangle=0,
\]
\begin{equation} \label{Noisecor}
\langle \xi_1({\bf r},t) \xi_1({\bf r'},t') \rangle=\langle
\xi_2({\bf r},t) \xi_2({\bf r'},t') \rangle=2\zeta \delta^d({\bf
r}-{\bf r'})\delta(t-t'),
\end{equation}
\[
\langle \xi_1({\bf r},t) \xi_2({\bf r'},t') \rangle=0.
\]

Similar to the expansion in the section III, we expand
\eqref{activ-inhabit} about the steady state, ${\bf u}({\bf
r},t)={\bf X}-{\bf X}_0$, where for the two-component system
${\bf u}({\bf r},t)=\left( \begin{array}{cc} u_1({\bf r},t)\\
u_2({\bf r},t) \end{array} \right)$. By choosing the linear part of
the expanded equations and transforming to Fourier space, the system
of equations \eqref{activ-inhabit} takes the form
\begin{equation} \label{Noise}
i\omega\left( {{\begin{array}{cc}
 \tilde{u}_1({\bf k},\omega)  \\
 \tilde{u}_2({\bf k},\omega) \\
\end{array}}} \right)-
\left( {{\begin{array}{cc}
 L_{11}-D_{\alpha1} k^\alpha & L_{12} \\
 L_{21} & L_{22}-D_{\alpha2}k^\alpha \\
\end{array}}}\right)
 \left( {{\begin{array}{cc}
 \tilde{u}_1({\bf k},\omega)  \\
 \tilde{u}_2({\bf k},\omega) \\
\end{array}}} \right)=\left( {{\begin{array}{cc}
 \tilde{\xi}_1({\bf k},\omega)  \\
 \tilde{\xi}_2({\bf k},\omega) \\
\end{array}}} \right),
\end{equation}
where $L_{ij}$'s are the components of the Jacobian matrix. The $2\times 2$ matrix in \eqref{Noise}
\[
{\cal L}=\left( {{\begin{array}{cc}
        L_{11}-D_{\alpha1} k^\alpha & L_{12} \\
        L_{21} & L_{22}-D_{\alpha2}k^\alpha \\
        \end{array}}}\right),
\]
contains the information about the critical behavior of the system.

To proceed towards studying the system near the Turing instability
we first need to find the eigenvalues and eigenvectors of $\cal L$
at the critical point as well as an analytic expression for its
eigenvalues close to the critical point. Therefore in the subsection
A we calculate these quantities first and find out how they vary in
the presence of superdiffusion. Then we use them to study the
behavior of the system in subsection B.

\subsection{The eigenvalue of the slow mode near the criticality}

The eigenvalues of $\cal L$ can be derived from the characteristic equation
\begin{equation} \label{characteristic}
\lambda^2+g\lambda+h=0,
\end{equation}
where
\begin{equation} \label{g}
g=(D_{\alpha1}+D_{\alpha2})k^\alpha-L_{11}-L_{22},
\end{equation}
\begin{equation} \label{h}
h=(L_{11}-D_{\alpha1} k^\alpha)(L_{22}-D_{\alpha2} k^\alpha)-L_{12}
L_{21}.
\end{equation}
$g$ and $h$ are functions of $k^\alpha$ and through $L$'s they also
depend on the bifurcation parameter $\mu$. Conventionally a parameter $B$ is defined in such a way that $\mu = 0$ corresponds to a bifurcation at $B=B_c$. The steady state
is linearly stable if and only if both $g$ and $h$ are non-negative
for all $k$. Clearly, this stability condition can be violated in
either of the following two ways:

1) $g$ vanishes for some $k$, but otherwise $g$ and $h$ remain
positive for all $k$. This condition together with ${\partial
g}/{\partial k}|_{k=k_c}=0$, which determines the critical $k_c$,
characterize a Hopf bifurcation that was discussed in the previous
section. Note that in this case $k_c=0$ (see \eqref{g}).

2) $h$ vanishes for some $k$, but otherwise $g$ and $h$ remain
positive for all $k$. This condition together with ${\partial
h}/{\partial k}|_{k=k_c}=0$ is adequate to determine the critical
values, $k_c$ and $B_c$, for a Turing bifurcation. For these critical
values, \eqref{characteristic} has two solutions
\[
\lambda_{s}^c=0, \;\;\;\; \lambda_{f}^c=-g(k_c^\alpha,B_c)\equiv
-g_c,
\]
where the subscripts $s$ and $f$ stand respectively for slow and
fast modes and $\lambda_s^c=\lambda_s(k_c^\alpha,B_c)$
($\lambda_f^c=\lambda_f(k_c^\alpha,B_c)$). The mode with eigenvalue
$\lambda_{s}^c=0$ is the mode which emerges and become macroscopic
at the critical point. So in the instability investigations this
mode becomes important. Let us calculate $\lambda_s(k,B)$ near the
critical point. The characteristic equation \eqref{characteristic}
has two general solutions
\begin{equation} \label{root}
\lambda=-\frac{1}{2}g\pm \frac{1}{2}\sqrt{(g^2-4h)}.
\end{equation}
On the other hand, the double Taylor expansion series of $\lambda_s(k,B)$ about the critical point $(k^\alpha_{c},B_c)$ is
\begin{equation} \nonumber
\lambda_s(k,B)=\frac{\partial \lambda}{\partial B}\bigg{|}_c
(B-B_c)+\frac{\partial \lambda}{\partial (k^\alpha)}\bigg{|}_c
(k^\alpha-k_c^\alpha)+\frac{1}{2}\frac{\partial^2 \lambda}{\partial
B^2}\bigg{|}_c (B-B_c)^2+\frac{1}{2}\frac{\partial^2 \lambda}{\partial
(k^\alpha)^2}\bigg{|}_c (k^\alpha-k_c^\alpha)^2+... .
\end{equation}
Therefore, using \eqref{root}, $\lambda_s$ can be written as
\begin{equation} \label{lambda}
\lambda_s(k,B)\approx-\frac{1}{g_c}\bigg{[}\frac{\partial
h}{\partial B}\bigg{|}_c (B-B_c)+D_{\alpha1}D_{\alpha2} (k^\alpha-k_c^\alpha)^2\bigg{]}.
\end{equation}
\eqref{lambda} shows how the eigenvalue of the slow mode near the
critical point depends on the superdiffusion exponent. In addition,
the matrix $\mathcal{L}$ at the critical point has the left eigenvector
corresponding to the eigenvalue $\lambda_s^c$
\begin{equation} \label{left2}
{\bf \mathcal{U}_{Ls}}(k_c,B_c)=C(U_1\;\;\;U_2),
\end{equation}
where $U_1=\frac{D_{\alpha 2} k_c^\alpha
-L_{22}(B_c)}{L_{12}(B_c)}$, $U_2=1$, and $C$ is a constant
that can be determined using the normalization condition and is not
important here.

Equipped with this information, \eqref{lambda} and \eqref{left2}, in
the next subsection we will study the dynamics of the slow mode as
well as the fluctuations close to the Turing instability.

\subsection{Generalized free energy}

Ginzburg-Landau theory has proven to be a very useful tool for the
analysis of non-equilibrium structures
\cite{Reichl,Walgraef2,Graham,Swift}. The reason is that in the
neighborhood of the critical point some modes exist on much slower
time scales than the other modes. Their behavior can be isolated and
analyzed independently of the other degrees of freedom. In this
subsection we will show how this may be done for the onset of Turing
patterns in a reaction-superdiffusion system. Actually we isolate
the slow mode from the equation of motion \eqref{Noise} to study its
dynamics. Also we obtain the correlation function of this mode near
the critical point to study the fluctuations close to the Turing
instability.

Let us isolate the critical eigenmode from the equation of motion. If we multiply \eqref{Noise} by the left eigenvector \eqref{left2} we find
\begin{equation} \label{isol}
(i\omega-\lambda_s(k^\alpha,B))\tilde{\varphi}({\bf
k},\omega)=\tilde{\eta}({\bf k},\omega),
\end{equation}
where we have approximated the left eigenvector of $\mathcal{L}$ by
its value at the critical point, \eqref{left2}, and $\lambda_s(k,B)$ is
given by \eqref{lambda}. $\tilde{\varphi}({\bf k},\omega)$ is the
amplitude of the critical eigenmode
\cite{Reichl,Walgraef2,Graham,Swift} that is a linear combination of
$\tilde{u}_1$ and $\tilde{u}_2$
\begin{equation}\label{linearphi}
\tilde{\varphi}({\bf k},\omega)=U_1\tilde{u}_1({\bf
k},\omega)+\tilde{u}_2({\bf k},\omega),
\end{equation}
and
\begin{equation} \label{etak}
\tilde{\eta}({\bf k},\omega)=U_1\tilde{\xi}_1({\bf
k},\omega)+\tilde{\xi}_2({\bf k},\omega),
\end{equation}
is the noise it experiences.
To study the fluctuations near the critical point, we obtain the correlation function for the critical eigenmode as
\begin{equation}\label{corr}
\langle\tilde{\varphi}({\bf k},\omega)\tilde{\varphi}({\bf
k'},\omega')\rangle=\frac{\langle\tilde{\eta}({\bf
k},\omega)\tilde{\eta}({\bf
k'},\omega')\rangle}{(i\omega-\lambda_{s}(k,B))(i\omega'-\lambda_{s}(k',B))}.
\end{equation}
Making use of equations \eqref{Noisecor} and \eqref{etak}, one can
find the following expression for the noise correlation function
\begin{equation}\label{noisecorr}
\langle\tilde{\eta}({\bf k},\omega)\tilde{\eta}({\bf
k'},\omega')\rangle=2(2\pi)^{d_s+1}\zeta(1+U_1^2)\delta^{d_s}({\bf k}+{\bf
k'})\delta(\omega+\omega'),
\end{equation}
and therefore \eqref{corr} becomes
\begin{equation} \label{Phi1}
\langle\tilde{\varphi}({\bf k},\omega)\tilde{\varphi}({\bf
k'},\omega')\rangle=\frac{2\zeta(2\pi)^{d_s+1}(1+U_1^2)\delta^{d_s}({\bf
k}+{\bf
k'})\delta(\omega+\omega')}{\omega^2+\lambda^2_{s}(k^\alpha,B)},
\end{equation}
where $d_s$ is the dimension of space.
On the other hand, by inverse Fourier transform in time of $\langle\tilde{\varphi}({\bf k},\omega)\tilde{\varphi}({\bf
k'},\omega')\rangle$, after some calculations and change of variables according to \cite{Reichl}, we can generally write
\begin{equation} \label{Phi2}
\langle\tilde{\varphi}({\bf k},\omega)\tilde{\varphi}({\bf
    k'},\omega')\rangle=(2\pi)^{d_s+1} \delta^{d_s}({\bf k}+{\bf
k'})\delta(\omega+\omega')\int_{-\infty}^{+\infty} dT e^{-i\omega T}
\langle \varphi({\bf k},T) \varphi({\bf -k},0) \rangle,
\end{equation}
where $T=t'-t$. Making comparison between \eqref{Phi1} and \eqref{Phi2} shows that
\begin{equation} \label{comp}
\int_{-\infty}^{+\infty} dt e^{-i\omega
t}\langle\tilde{\varphi}({\bf k},t)\tilde{\varphi}({\bf
-k},0)\rangle=\frac{2\zeta(U_1^2+1)}{\omega^2+\lambda_{s}^2}.
\end{equation}
The inverse Fourier transform in time of \eqref{comp} results in
\begin{equation}
\langle\tilde{\varphi}({\bf k},t)\tilde{\varphi}({\bf
-k},0)\rangle={\frac{\zeta (1+U_1^2)}{\pi}}\int_{-\infty}^{+\infty} d\omega
e^{i\omega t}(\frac{1}{\omega^2+\lambda_{s}^2}),
\end{equation}
and after integration, we find
\begin{equation} \label{correlation}
\langle\tilde{\varphi}({\bf k},t)\tilde{\varphi}({\bf
-k},0)\rangle=\zeta
(1+U_1^2)\frac{e^{-\lambda_{s}t}}{\lambda_s}.
\end{equation}
According to \eqref{correlation}, the relaxation time is
$\lambda_s^{-1}(k,B)$ where $\lambda_s$ is given by \eqref{lambda}.
There are two noteworthy points as we approach the critical point.
First, the fluctuations in the critical eigenmode takes longer and
longer to die away which is the critical slowing down. Second, the
correlation function \eqref{correlation} begins to diverge due to
the dependence on $\lambda_s$ in the denominator which resembles an
equilibrium phase transitions.

It can be easily verified that the equation of motion for the
critical eigenmode \eqref{isol} can be written in the form of a
time-dependent Ginzburg-Landau equation
\begin{equation}\label{GL}
\frac{\partial \tilde{\varphi}({\bf k},t)}{\partial t}=\frac{\delta
F}{\delta \tilde{\varphi}(-{\bf k},t)}+\tilde{\eta}({\bf k},t),
\end{equation}
in which the Ginzburg-Landau free energy, $F$, is given by
\begin{equation}\label{free}
\begin{split}
F=&\frac{1}{2}\int d {\bf k} \lambda_s(k,  B)|\tilde{\varphi}({\bf k},t)|^2 \\
=&-\frac{1}{2g_c}\int d{\bf k}
\bigg{(}\frac{\partial
    h}{\partial B}\bigg{|}_c (B-B_c)+D_{\alpha1}D_{\alpha2} (k^\alpha-k_c^\alpha)^2\bigg{)}|\tilde{\varphi}({\bf k},t)|^2,
\end{split}
\end{equation}
where the noise correlator is \eqref{noisecorr}. The correlation
function \eqref{correlation} can also be directly calculated from
the time-dependent Ginzburg-Landau equation \eqref{GL}. Note that in order to
write \eqref{free} one needs an expression for $\lambda_s$ which, for
a two-component system, is given analytically by \eqref{lambda}. \eqref{free} also indicates that the free energy depends on
superdiffusion exponent.

Therefore as we claimed at the beginning of this section, there is a
generalized thermodynamic potential (Ginzburg-Landau free energy)
that can describe the dynamics and fluctuations of the
reaction-superdiffusion system near the Turing instability.

\section{Brusselator model with superdiffusion}

Our plan in this section is to apply what we did in the previous
sections to a typical reaction-superdiffusion system called
Brusselator model. In the first subsection, A, we analyze the
stability of the system and then in subsections B and C we
respectively investigate the behavior of this system in the vicinity
of the Hopf and Turing bifurcations.

The Brusselator is one of the simplest models of a nonlinear
chemical system for which the relative concentration of the
constituents can oscillate in time (chemical clock) or has
stationary concentration patterns (chemical crystals) \cite{Reichl}.
The Brusselator model is a two-component activator-inhibitor system that
in the case of Levy flight takes the form
\begin{equation} \label{FracBrus}
\left\{ \begin{array}{rcl} &~&
{\frac{\partial X_1}{\partial t}=D_{\alpha1} \nabla^\alpha X_1+A-(B+1)X_1+X_1^2X_2}\\
&~& \frac{\partial X_2}{\partial t}=D_{\alpha2} \nabla^\alpha
X_2+BX_1-X_1^2X_2
\end{array}\right.,
\end{equation}
where $X_1$ and $X_2$ are chemical concentrations that can vary in
space and time and $A$ and $B$ are constants.

\subsection{Linear stability analysis}

To analyze the stability of the Brusselator system we first find the
steady state solution of \eqref{FracBrus} which is $(A,B/A)$. Then
we consider perturbations about the steady state as
\begin{equation} \nonumber
X_1({\bf r},t)=A+u_1({\bf r},t), \;\;\;\;\;\;\; X_2({\bf
r},t)=\frac{B}{A}+u_2({\bf r},t),
\end{equation}
and put them in \eqref{FracBrus}. The linearized equations of perturbations will be
\begin{equation}  \label{BrusLin}
\left\{ \begin{array}{rcl} &~&
{\frac{\partial u_1}{\partial t}=(B-1+D_{\alpha1} \nabla^\alpha)u_1+A^2u_2}\\
&~& \frac{\partial u_2}{\partial t}=-Bu_1+(-A^2+D_{\alpha2}
\nabla^\alpha) u_2
\end{array}\right..
\end{equation}
Substituting the normal mode solution
\[
\left( {{\begin{array}{cc}
 u_1  \\
 u_2 \\
\end{array}}} \right)= \left( {{\begin{array}{cc}
 a \\
 b \\
\end{array}}} \right) \rm exp(\lambda t+i{\bf k}.{\bf r}),
\]
into \eqref{BrusLin} results in
\begin{equation} \nonumber
\left( {{\begin{array}{cc}
        \lambda -B+1+D_{\alpha1} k^\alpha & -A^2 \\
        B & \lambda+A^2+D_{\alpha2} k^\alpha \\
        \end{array}}}   \right)
\left( {{\begin{array}{cc}
        a  \\
        b \\
        \end{array}}} \right)=0.
\label{3}
\end{equation}
So the characteristic equation becomes
\begin{equation} \nonumber
\lambda^2+g(k) \lambda+h(k)=0,
\end{equation}
where $k\equiv |\bf{k}|$, $g(k)=A^2-B+1+(D_{\alpha1} +D_{\alpha2})
k^\alpha$ and $h(k)=A^2+(A^2 D_{\alpha1}+ D_{\alpha2}- B
D_{\alpha2}) k^\alpha +D_{\alpha1}D_{\alpha2}k^{2\alpha}$ according
to \eqref{g}, \eqref{h}.

The modes that emerge in a system by Turing instability have no
oscillation in time but only in space. These patterns are called
chemical crystals in the Brusselator \cite{Reichl}. According to our
discussion in the section IV one can obtain the Turing instability
condition for the Brusselator as
\begin{equation} \nonumber
B(k)>1+\frac{D_{\alpha 1}}{D_{\alpha 2}}A^2+\frac{A^2}{D_{\alpha
2}k^\alpha}+D_{\alpha 1}k^\alpha,
 \label{3}
\end{equation}
which depends on $\alpha$. Thus, superdiffusion affects the
instability condition. The critical values $k_{cT}$ and $B_{cT}$ for Turing instability can
be found as
\begin{equation}\label{Turingcritical}
 B_{cT}={\bigg (}1+A \sqrt{\frac{D_{\alpha 1}}{D_{\alpha 2}}}{\bigg )}^2,\;\;\;\;\; \;\;\;\;\; k_{cT}={\bigg (}\frac{A}{\sqrt{D_{\alpha 1}D_{\alpha
 2}}}{\bigg )}^{1/\alpha}.
\end{equation}
\eqref{Turingcritical} shows that the critical parameters change with $\alpha$. Variation of the critical wave number of Turing patterns, $k_{cT}$, implies the change of the size of the emerged Turing patterns. So, depending on the parameters of the system, according to  \eqref{Turingcritical} superdiffusion can make the size of the patterns larger or smaller.

For a Hopf bifurcation in a reaction-superdiffusion system the
emerged modes have an oscillatory behavior in time. Such systems are
called chemical clocks in the Brusselator \cite{Reichl}. Again
according to the instability discussion in section IV the Hopf
instability condition for the Brusselator is
\begin{equation}\label{Hopfcondition}
 B(k)> 1+A^2+(D_{\alpha1}+D_{\alpha2})k^\alpha,
\end{equation}
and the critical values of $k$ and $B$ in this case are
\begin{equation}\label{Hopfcritical}
k_{cH}=0, \;\;\;\;\;\;\;\;\;\;  B_{cH}=1+A^2.
\end{equation}
Note that superdiffusion changes the instability condition for a
Hopf bifurcation, \eqref{Hopfcondition}, but it does not change the
critical parameters, \eqref{Hopfcritical}.

\subsection{Hopf instability in Brusselator}

To study the Hopf instability in the Brusselator model we start from
\eqref{Taylor} in the section III and follow the presented method to finally find the FCGLE governing the system.

General equation \eqref{Taylor} which governs the dynamics of the
perturbations reduces to the following equation for the Brusselator
as a two-component model
\begin{equation} \nonumber
\left( {{\begin{array}{cc}
 \dot{u}_1  \\
 \dot{u}_2 \\
\end{array}}} \right)=
\left( {{\begin{array}{cc}
 B-1 &\;\;\; A^2 \\
 -B & \;\;\; -A^2 \\
\end{array}}}\right)
 \left( {{\begin{array}{cc}
 u_1 \\
 u_2 \\
\end{array}}} \right)+ \mathcal{D}_\alpha \nabla^\alpha \left( {{\begin{array}{cc}
 u_1 \\
 u_2 \\
\end{array}}} \right)+ \left( {{\begin{array}{cc}
 \beta  \\
 -\beta \\
\end{array}}} \right),
\end{equation}
where $\beta$ in the non-linear term is  $\beta=\frac{B}{A}u_1^2+2Au_1u_2+u_1^2u_2$ and $\mathcal{D}_\alpha=\left( {{\begin{array}{cc}
D_{1\alpha} & 0 \\
0 & D_{2\alpha} \\
\end{array}}} \right)$.
Since the system is supposed to experience a Hopf instability,
according to \eqref{Hopfcritical} the zeroth order of the matrix $L$
($L$ at critical point) will be
\begin{equation} \nonumber
L^{(0)}= \left( {{\begin{array}{cc}
A^2 & \;\; A^2 \\
-(1+A^2) & \;\; -A^2 \\
\end{array}}} \right),
\end{equation}
with the eigenvalues $\lambda^{(0)}=\pm iA$. Considering
$\mu=(B-B_c)/B_c$, the first order of $L$ (see \eqref{expand}) is
\begin{equation} \nonumber
L^{(1)}= (1+A^2) \left( {{\begin{array}{cc}
 1 \;\;& 0 \\
 -1 \;\;& 0 \\
\end{array}}} \right).
\end{equation}

Following the reductive perturbation theory of the section III, one
can find the FCGLE, \eqref{CGLEF}, for the Brusselator. The coefficients $\lambda^{(1)}$, $d$ and $g$ are found from Eq.
\eqref{CGLEcoeff} to be
\begin{equation} \nonumber
\lambda^{(1)}=\frac{1+A^2}{2},
\end{equation}
\begin{equation} \nonumber
d=\frac{1}{2}[D_{1\alpha}+D_{2\alpha}-iA(D_{1\alpha}-D_{2\alpha})],
\end{equation}
\begin{equation} \nonumber
g=\frac{1}{2}(\frac{2+A^2}{A^2}+i\frac{4-7A^2+4A^4}{3A^3}).
\end{equation}
The obtained FCGLE can be written in the standard form,
\eqref{CGLERS}, with the parameters
\begin{equation} \nonumber
c_1=-A
\frac{D_{1\alpha}-D_{2\alpha}}{D_{1\alpha}+D_{2\alpha}}\;\;\;\;\;
\rm and \;\;\;\;\ c_2=\frac{4-7A^2+4A^4}{3A(2+A^2)}.
\end{equation}
As can be seen, superdiffusion can change the properties of the solution
through the coefficient $c_1$ as well as changing the order of the
derivative ($\nabla^\alpha$).

\subsection{Turing instability in the Brusselator}

In this subsection we apply the general topics about the behavior of
a reaction-superdiffusion system near Turing instability, presented
in the section IV, to the Brusselator model and finally find the
Ginzburg-Landau free energy which governs the behavior of this
system.

The general linearized equation \eqref{Noise} in the case of the Brusellator takes the form
\begin{equation}   \label{BrusNoise}
i\omega\left( {{\begin{array}{cc}
 \tilde{u}_1({\bf k},\omega)  \\
 \tilde{u}_2({\bf k},\omega) \\
\end{array}}} \right)-
\left( {{\begin{array}{cc}
 B-1-D_{\alpha1} k^\alpha & A^2 \\
 -B & -A^2-D_{\alpha2}k^\alpha \\
\end{array}}}\right)
 \left( {{\begin{array}{cc}
 \tilde{u}_1({\bf k},\omega)  \\
 \tilde{u}_2({\bf k},\omega) \\
\end{array}}} \right)=\left( {{\begin{array}{cc}
 \tilde{\xi}_1({\bf k},\omega)  \\
 \tilde{\xi}_2({\bf k},\omega) \\
\end{array}}} \right),
\end{equation}
where here ${\xi}_1({\bf r},t)$ and ${\xi}_2({\bf r},t)$ are random
chemical noises and $\tilde{u}_1({\bf k},\omega)$ and
$\tilde{u}_2({\bf k},\omega)$ represent the Fourier transform of
concentration fluctuations. The eigenvalues of the matrix ${\cal L}$
in \eqref{BrusNoise} at critical point can be obtained using the
critical values $k_{cT}$ and $B_{cT}$ in \eqref{Turingcritical}
\begin{equation} \nonumber
\lambda_{s}^c=0,\;\;\;\lambda_{f}^c=A\left(\sqrt{\frac{D_{\alpha1}}{D_{\alpha2}}}+A\frac{D_{\alpha1}}{D_{\alpha2}}-A-\sqrt{\frac{D_{\alpha2}}{D_{\alpha1}}}\right)\equiv
-g_{cB},
\end{equation}
where $g_{cB}$ stands for $g_c$ in Brusselator.
The left eigenvector corresponding to the slow mode at the critical point is
\begin{equation} \nonumber
{\bf \mathcal{U}_{Ls}}(k_c,B_c)\propto
(1+\frac{1}{D_{\alpha1}k^\alpha_c}\;\;\;\;1).
 \label{3}
\end{equation}
Also, regarding \eqref{lambda}, the eigenvalue of the slow mode in the vicinity of the critical point is given by
\begin{equation} \label{lamBrus}
\lambda_{s}(k,B)\approx \frac{1}{g_{cB}}[
A\sqrt{\frac{D_{\alpha2}}{D_{\alpha1}}}(B-B_c)-D_{\alpha1}D_{\alpha2}(k^\alpha-k^\alpha_c)^2].
\end{equation}
Putting everything together, the amplitude of the critical eigenmode
for the Brusselator, according to \eqref{linearphi}, will be
\begin{equation}\nonumber
\tilde{\varphi}({\bf
k},\omega)=(1+\frac{1}{D_{\alpha1}k^\alpha_c})\tilde{u}_1({\bf
k},\omega)+\tilde{u}_2({\bf k},\omega),
\end{equation}
and it experiences the following chemical noise
\begin{equation} \nonumber
\tilde{\eta}({\bf
k},\omega)=(1+\frac{1}{D_{\alpha1}k^\alpha_c})\tilde{\xi}_1({\bf
k},\omega)+\tilde{\xi}_2({\bf k},\omega),
\end{equation}
with the correlator
\begin{equation}\label{noisecorrBruss}
\langle\tilde{\eta}({\bf k},\omega)\tilde{\eta}({\bf
    k'},\omega')\rangle=
2(2\pi)^{d_s+1}\zeta[1+(1+\frac{1}{k_c^\alpha
D_{\alpha1}})^2]\delta^{d_s}({\bf k}+{\bf
k'})\delta(\omega+\omega').
\end{equation}

The correlation function of the critical eigenmode, \eqref{correlation}, is
\begin{equation}\nonumber
\langle \tilde{\varphi}({\bf k},t) \tilde{\varphi}(-{\bf
k},0))\rangle=[1+(1+\frac{1}{k_c^\alpha
D_{\alpha1}})^2]\frac{e^{-\lambda_{s}t}}{\lambda_{s}},
\end{equation}
where $\lambda_{s}$ is given by \eqref{lamBrus}. As discussed at the
end of the section IV, the dynamics of the fluctuations near the
Turing instability is obtained from a time-dependent Ginzburg-Landau
equation, \eqref{GL}, in which the Ginzburg-Landau free energy, here
for the Brusselator, is given by
\begin{equation}\nonumber
F=-\frac{1}{2g_{cB}}\int d{\bf k}
\bigg{(}-A\sqrt{\frac{D_{\alpha2}}{D_{\alpha1}}} (B-B_c)+D_{\alpha1}D_{\alpha2} (k^\alpha-k_c^\alpha)^2\bigg{)}|\tilde{\varphi}({\bf k},t)|^2,
\end{equation}
with the correlator \eqref{noisecorrBruss} for the noise term.

\section{Conclusion and discussion}

In this paper we studied the effect of Levy superdiffusion on the
instability of a general reaction-superdiffusion system for two
possible kinds of instabilities, Hopf and Turing. Superdiffusion can
be considered in reaction-diffusion systems using the powerful tool
of fractional calculus. Utilizing the reductive perturbation theory
we showed that a fractional complex Ginzberg-Landau equation governs
the amplitude of the critical mode near a Hopf bifurcation for a
general \textit{n-component} reaction-superdiffusion system.
Numerical simulations based on pseudospectral method were performed
to study FCGLE solutions in two dimensions. We showed that as one
decreases the fractional order, larger size cellular structures
appear in frozen states; density of defects reduces in the defect
turbulence regime; and the amplitude of fluctuations increases in
the phase turbulence regime. In addition we deduced that
superdiffusion can create and annihilate spiral patterns and can
make a transition from one regime to another. We also studied FRGLE
as a limiting case of FCGLE and observed that the size of the
stationary patterns increases in the presence of superdiffusion.

As the main part of our work we studied the behavior of a general
reaction-superdiffusion system in the vicinity of a Turing
instability and derived its dependence on the superdiffusion
exponent. It was shown that reaction-superdiffusion systems close to
a Turing instability are equivalent to a time-dependent
Ginzburg-Landau model and the corresponding free energy was
introduced which depends on superdiffusion exponent. Including a local white noise in the system we obtained
the correlation function for the critical eigenmode in the linear
approximation. This correlation function diverges at the critical
point which is analogous to equilibrium phase transitions. The
correlation function can also be directly obtained from the
presented time-dependent Ginzburg-Landau equation. In general, the
introduced generalized free energy governs the stability, dynamics
and the fluctuations of reaction-superdiffusion systems near the
Turing bifurcation.

We considered the Brusselator model in the presence of
superdiffusion as a typical example of a reaction-diffusion system.
The FCGLE and the generalized free energy respectively for the case
of Hopf and Turing instabilities were derived for the Brusselator.
In addition, linear stability analysis of the Brusselator indicated
changes in the instability conditions in the presence of
superdiffusion.

As a conclusion, superdiffusion introduces a new parameter that
changes the properties of the solutions. Therefore superdiffusion
can be exploited to control and change the dynamical properties of a reaction-diffusion
system.

\begin{acknowledgments}
The authors are grateful to the referees for their constructive comments. One of the authors, R. T, would like to thank G. R. Jafari for useful discussions.
\end{acknowledgments}

\end{document}